\documentclass[10pt,journal,compsoc]{IEEEtran}
\ifCLASSOPTIONcompsoc
  \usepackage[nocompress]{cite}
\else
  \usepackage{cite}
\fi
\usepackage{amsfonts,amssymb,amsmath,amsthm,array}
\usepackage[ruled,linesnumbered]{algorithm2e}  
\usepackage{bm,bbding,booktabs}
\usepackage[english]{babel}
\usepackage{calc,cases}
\usepackage{diagbox}
\usepackage{enumitem}
\usepackage{float}
\usepackage{graphicx}
\usepackage[colorlinks=true, allcolors=blue]{hyperref}
\usepackage{mathrsfs,multirow}
\usepackage{tabularx,tikz,textcomp}
\usepackage{url}
\usepackage{verbatim}
\usepackage{pifont}
\usepackage{subfigure,stfloats}
\usepackage{xcolor}

\newtheorem{theorem}{Theorem}
\newtheorem{lemma}{Lemma}
\newtheorem{assumption}{Assumption}
 
\graphicspath{{figures/}}

\def\BibTeX{{\rm B\kern-.05em{\sc i\kern-.025em b}\kern-.08em
    T\kern-.1667em\lower.7ex\hbox{E}\kern-.125emX}}

\ifCLASSINFOpdf
\else
\fi
\begin{document}

\title{{LiteChain}: A Lightweight Blockchain for Verifiable and Scalable Federated Learning in Massive Edge Networks}
\author{Handi~Chen,
        Rui~Zhou,
        Yun-Hin~Chan,
        Zhihan~Jiang,~\IEEEmembership{Graduate Student Member,~IEEE,} 
        Xianhao~Chen,~\IEEEmembership{Member,~IEEE,} 
        and~Edith~C.~H.~Ngai,~\IEEEmembership{Senior Member,~IEEE}% <-this % stops a space
\IEEEcompsocitemizethanks{\IEEEcompsocthanksitem H. Chen, R. Zhou, Y. Chan, Z. Jiang, X. Chen and E. Ngai are with the Department of Electrical and Electronic Engineering, The University of Hong Kong, Hong Kong 999077, China.\protect\\
(E-mail: \{hdchen; zackery; chanyunhin; zhjiang\}@connect.hku.hk; \{xchen; chngai\}@eee.hku.hk.) (Corresponding author: Edith C. H. Ngai.)
% \thanks{Manuscript received April 19, 2005; revised August 26, 2015.}
}}

% \markboth{IEEE Transactions on mobile computing}%
% {Shell \MakeLowercase{\textit{et al.}}: Bare Demo of IEEEtran.cls for Computer Society Journals}

\IEEEtitleabstractindextext{%
\begin{abstract}
Leveraging blockchain in Federated Learning (FL) emerges as a new paradigm for secure collaborative learning on Massive Edge Networks (MENs). As the scale of MENs increases, it becomes more difficult to implement and manage a blockchain among edge devices due to complex communication topologies, heterogeneous computation capabilities, and limited storage capacities. Moreover, the lack of a standard metric for blockchain security becomes a significant issue. To address these challenges, we propose a lightweight blockchain for verifiable and scalable FL, namely LiteChain, to provide efficient and secure services in MENs. Specifically, we develop a distributed clustering algorithm to reorganize MENs into a two-level structure to improve communication and computing efficiency under security requirements. Moreover, we introduce a Comprehensive Byzantine Fault Tolerance (CBFT) consensus mechanism and a secure update mechanism to ensure the security of model transactions through LiteChain. Our experiments based on Hyperledger Fabric demonstrate that LiteChain presents the lowest end-to-end latency and on-chain storage overheads across various network scales, outperforming the other two benchmarks. In addition, LiteChain exhibits a high level of robustness against replay and data poisoning attacks.
\end{abstract}

}

\maketitle

\IEEEdisplaynontitleabstractindextext
\IEEEpeerreviewmaketitle

\IEEEraisesectionheading{\section{Introduction}\label{sec:introduction}}
\IEEEPARstart{T}{he} evolution of wireless communication and sensing complicates the network topology among heterogeneous and resource-constrained edge devices, such as gateways, mobile phones, and other computing devices. The increasing number of edge devices contributes to the rapid expansion of edge networks, leading to Massive Edge Networks (MENs) \cite{guo2021enabling}. The number of Internet of Things (IoTs) connected devices is expected to exceed 22 billion by 2025 \cite{Jouhari2023survey}. 
Managing such a vast number of devices in a centralized manner becomes virtually impractical, making it necessary to decentralize MENs. 
Furthermore, devices in MENs engage in more frequent and complex interactions, necessitating the implementation of more efficient authorization mechanisms to ensure both security and prompt response. To extract valuable information from raw data generated by massive edge devices in communication-driven MENs, machine learning applications facilitate the development of intelligence-driven MENs \cite{tang2022internet, liu2023mars3d, zhou2019edge}. However, the substantial data transmission requirements for centralized model training raise privacy concerns. In light of this, Federated Learning (FL), collaboratively training a shared model without private data exposure, becomes a pivotal enabler for MENs intelligence by reducing the risk of data leakage and abuse.  

However, the assumptions of trustworthy central servers and clients in traditional FL studies are often unrealistic in the real world \cite{chen2022blockchain}. Malicious devices may attack model training to affect the convergence process adversely. Prior work has proposed various defense mechanisms for verifiable global model from poisoned data in aggregation, such as Krum/Multi-Krum \cite{blanchard2017machine}, trimmed-mean-based gradient descent \cite{melis2019exploiting}, auto-weighted robust federated learning \cite{li2022auto}, and ClippedClustering \cite{li2023experimental}. Nevertheless, the framework is still vulnerable to data breaches and malicious tampering during data transmission when publicly accessible edge devices participate in training. To address these issues, a blockchain is often utilized to record gradients or model transactions for verifiable FL \cite{nguyen2021federated}. 

Due to its decentralization, immutability, and traceability characteristics, a blockchain provides a distributed and transparent ledger for information exchange for verifiable FL in IoTs, which has been demonstrated in numerous studies. Existing works on blockchains can be categorized into four types: private, public, consortium, and hybrid blockchains. Private blockchain-empowered FL systems \cite{rathore2019blockdeepnet, weng2019deepchain, warnat2021swarm} are designed for highly confidential IoTs controlled by a single trustworthy organization, which ensure verifiability and security highly effective but often at the cost of scalability. In contrast, public and consortium blockchains are more naturally compatible with MENs due to their ability to facilitate secure and transparent interactions among massive untrusted parties. Public blockchain-empowered FL systems, such as BlockFL \cite{kim2018device}, Biscotti \cite{shayan2020biscotti}, FL-Block \cite{qu2020decentralized}, and BAFL  \cite{feng2021bafl}, enable open participation in the consensus process, fostering a broad base of trust but also demand significant resources from the IoTs in which they are applied. Consortium blockchain-empowered FL systems facilitate consensus through a committee composed of a pre-selected group of nodes, providing more adaptable and efficient management for verifiable FL in collaborative IoTs, such as FabricFL \cite{mothukuri2021fabricfl} and B-FL \cite{yang2022trustworthy}. Hybrid blockchain-empowered FL systems combine elements from multiple types of blockchains to balance controlled access and efficiency for IoT applications that involve diverse stakeholders and requirements, such as BEFL \cite{jin2023lightweight}, VFChain \cite{peng2021vfchain}, PermiDAG \cite{lu2020blockchain}.

However, existing blockchains implemented in MENs result in high resource consumption and limited scalability. The complexity of coordinating and synchronizing diverse IoT devices within MENs, along with ensuring scalable and efficient data transactions while maintaining privacy and security across the blockchain, further complicates deployment and reduces operational efficiency. Moreover, lacking unified standards to measure security is also a challenge in evaluating diverse blockchain architectures. To summarize, there are four main challenges in the current blockchain-based solutions for FL in MENs: 1) high computation demands from maintaining blockchain consensus mechanisms, 2) network congestion resulting from high communication complexities, 3) the excessive burden of on-chain storage costs detrimental to sustainable training, and 4) system security lacking quantitative metrics for comparison. Therefore, there is a significant need to develop a lightweight blockchain framework to facilitate effective and efficient FL training in MENs.

To address these challenges, we design a lightweight blockchain framework for verifiable and scalable FL in MENs to reduce the training burden for devices from multiple dimensions. Before initiating the FL training task, a distributed clustering algorithm is used to reorganize the network topology into a hierarchical structure autonomously. The reorganized network consists of multiple clusters with elected committee members and is optimized by balancing network latency and consensus security. During training, all elected committee members construct a committee and take on dual roles: 1) aggregators executing intra-cluster and inter-cluster aggregations and synchronizing models within clusters, and 2) verifiers verifying the generated blocks. After a certain period, the blockchain initiates an update consensus mechanism to synchronize the model version and the committee membership. The contributions of the paper are highlighted as follows:
\begin{itemize}
    \item We propose a distributed clustering optimization algorithm to reorganize MENs into a communication-efficient hierarchical structure by balancing latency and {consensus security}. To minimize idle computation capacities, we adopt two-tier FL training that features intra-cluster and inter-cluster aggregation.
    \item For security, we propose a quantitative formula to measure the on-chain {consensus security}. Moreover, we employ a Comprehensive Byzantine Fault Tolerance (CBFT) consensus mechanism  to verify model quality and block integrity to secure model transactions. Additionally, our update consensus mechanism synchronizes models periodically to avoid power centralization.
    \item To ease the burdens of on-chain storage and data transmission, we introduce model identifiers to block storage. In addition, leveraging the new update consensus mechanism, LiteChain further reduces the redundancy of stale storage through periodic synchronization.
    \item We conduct extensive experiments over two benchmarks across 50 to 300 devices with and without attacks. Experimental results show that LiteChain achieves the lowest latency at the stopping criterion and is less prone to attacks under various network scales.
\end{itemize}

The rest of this paper is organized as follows: The related work is reviewed in section 2. Section 3 elaborates on two crucial objective models -- latency and {consensus security} models. The design of LiteChain is described in Section 4. Experiments are conducted and analyzed in Section 5, followed by the conclusion in Section 6.

\section{Related Work}
{In this section, we review representative studies of verifiable FL from the perspectives of centralized and decentralized training. Moreover, we provide a detailed analysis of blockchain-empowered verifiable FL studies {\cite{Nguyen2021FL}} categorized by the type of blockchain utilized, including public, private, consortium, and hybrid blockchains.}

\subsection{Verifiable FL}
{Verifiable FL enables participants to confirm the training process without disclosing the private data involved {\cite{xu2019verifynet}}. Common verifiable FL systems can be categorized based on the centralized and decentralized training process.}

{Centralized FL systems primarily involve two participants: clients and a central server {\cite{mcmahan2017communication}} {\cite{chan2024internal}}. The verification mechanisms implemented by the clients utilize various cryptographic techniques to ensure the correctness of the aggregation model received from the server. For instance, VERSA {\cite{hahn2021versa}} employs lightweight cryptographic primitives such as pseudorandom generators to ensure the integrity and authenticity of aggregated data across devices. VerifyNet {\cite{xu2019verifynet}} proposes a double masking protocol for encrypting local gradients and requires the cloud server to provide tamper-proof evidence of the correctness of the aggregation results.} 
{On the server side, verifiable FL focuses on verifying the integrity, authenticity, and correctness of the local training process while maintaining the confidentiality of the data used. TrustFL {\cite{zhang2020enabling}} verifies participants' contributions by adopting a ``commit-and-prove" mechanism where participants submit a commitment to indicate completion of training. This system randomly selects participants' digital signatures to authenticate each epoch. Moreover, the central server in the reputation-based method {\cite{wang2021reputation}} assigns weights to participants' model updates for aggregation based on the reputation scores calculated from historical test performance.}
 
{Decentralized FL involves multiple participants that directly communicate with each other without a central server. Each participant acts as both a trainer and an aggregator. Verification mechanisms are implemented to verify the correctness and timeliness of the received models before aggregating them with their models for next-round training. Chen \textit{et al.} {\cite{chen2021bdfl}} propose a Byzantine-Fault-Tolerance (BFT) verifiable decentralized FL based on HydRand protocol for the Internet of Vehicles (IoVs), named BDFL. BDFL enables participating vehicles to verify the correctness of the encrypted models using a publicly verifiable secret-sharing scheme.}
{Additionally, blockchain is a potent tool for verifiable decentralized FL due to its immutability, high transparency, and decentralized verification. We provide a detailed review of blockchain-empowered FL systems based on the categories of blockchain in Section 2.2.}

\subsection{Blockchain-Empowered FL}
\begin{table}[!t]
    \centering
    \renewcommand{\arraystretch}{1.2}
    \caption{Resource optimization analysis of existing blockchain-empowered FL systems from computation (Comp), communication (Comm), and storage resource perspectives. Here, \ding{51} and \ding{55} represent resources that are optimized and not optimized, respectively, compared to standard mechanism implementations. P2P and CS denote Peer-to-Peer and Client-Server FL training modes, respectively.}
    \label{tab:refe}
    \scalebox{0.85}{
    \begin{tabular}{>{\centering\arraybackslash}p{0.7cm}p{2.5cm}p{1.3cm}>{\centering\arraybackslash}p{0.9cm}>{\centering\arraybackslash}p{0.9cm}>{\centering\arraybackslash}p{0.9cm}}
    \toprule
    \textbf{Chain} & \textbf{Papers} & \textbf{FL}  & \textbf{Comp} & \textbf{Comm} & \textbf{Storage} \\
    \midrule
    \multirow{3}*{\rotatebox{90}{\textbf{Private}}}&BAFFLE\cite{ramanan2020baffle}  & P2P & \ding{51} &  \ding{55} & \ding{55}\\
    & BlockDeepNet\cite{rathore2019blockdeepnet} &CS & \ding{51} & \ding{55} &  \ding{55}\\
    & DeepChain\cite{weng2019deepchain} & P2P & \ding{51} & \ding{55} & \ding{55}\\
    
    \midrule
    \multirow{5}*{\rotatebox{90}{\textbf{Public}}}&Biscotti\cite{shayan2020biscotti}  & P2P & \ding{51} & \ding{55} & \ding{55}\\
    & FL-Block\cite{qu2020decentralized} & P2P & \ding{55} & \ding{55} &  \ding{51}\\
    & BAFL\cite{feng2021bafl} & P2P & \ding{51} &  \ding{55} &  \ding{55}\\
    & BRAFL\cite{ur2020towards} & P2P &  \ding{55} & \ding{51} & \ding{51} \\
    & PPBFL\cite{li2024ppbfl} & CS &  \ding{55} & \ding{55} & \ding{51}\\
    \midrule
    
    \multirow{4}*{\rotatebox{90}{\textbf{Consortium}}}&B-FL\cite{yang2022trustworthy} & CS & {{\ding{55}}} & {{\ding{51}}} & \ding{55} \\
    &BFLC\cite{li2020blockchain}  & P2P & {{\ding{55}}} & {{\ding{51}}} & {{\ding{51}}} \\
    &CBFL\cite{chen2020methodology}  & P2P & \ding{51} & \ding{51} & \ding{55} \\
    &{FabricFL}\cite{mothukuri2021fabricfl}  & CS& \ding{51} & \ding{55} & \ding{55} \\
    
    \midrule
    \multirow{6}*{\rotatebox{90}{\textbf{Hybrid}}}
    &{BE-DHFL}\cite{feng2021blockchain}  & CS & {{\ding{55}}} & {{\ding{55}}} & {{\ding{51}}} \\
    &HB\cite{chai2020hierarchical}  & CS & \ding{51} & \ding{55}  & \ding{55} \\
    &PermiDAG\cite{lu2020blockchain}  & CS & \ding{51} & {{\ding{51}}} & \ding{55} \\
    & BEFL\cite{jin2023lightweight} & P2P & \ding{55} & \ding{51} & \ding{51}\\ 
    \cmidrule{2-6}
    & \textbf{LiteChain} & CS \& P2P & \ding{51} & \ding{51}& \ding{51} \\
    \bottomrule
    \end{tabular}}
\end{table}

{Some representative blockchain-empowered FL frameworks and their resource optimization analysis on computation, communication, and storage are summarized in Table {\ref{tab:refe}}. {\ding{51} and \ding{55}} represent 
if a resource consumption has been reduced using the proposed blockchains compared to standard mechanisms. In Table {\ref{tab:refe}}, the ``FL" column indicates the FL training modes, including Peer-to-Peer (P2P) and Client-Server (CS) training modes.}

\subsubsection{Private Blockchain-Empowered FL}
Private blockchains are invite-only networks managed by an authoritative entity. All activities require permission through {verification} mechanisms \cite{nguyen2021federated}. Rathore \textit{et al.} \cite{rathore2019blockdeepnet} develop BlockDeepNet, {a private blockchain-based collaborative CS FL system without verification mechanism for big data analysis in 5G-enhanced IoT.} 
% BlockDeepNet utilizes {verification}-free private blockchain to provide secure and reliable exchanges of model updates on authorized IoT devices for Client-Server (CS) FL. 
% However, private blockchains are vulnerable to potential fraudulent attacks without consensus mechanisms. 
DeepChain \cite{weng2019deepchain} and BAFFLE \cite{ramanan2020baffle} are proposed based on proof-of-authority and blockwise-BA consensus mechanisms to maintain the security of model interactions during FL training, respectively. 
% This authorization-relied framework presents high practicality in privacy-sensitive scenarios. 
Warnat \textit{et al.} \cite{warnat2021swarm} {propose swarm learning leveraging private blockchain for secure FL on private information-sensitive healthcare networks to create disease classifiers. }
% Swarm learning enhances precision medicine with distributed models instead of private data to maintain local confidentiality. 

{Overall, given the smaller scale of private blockchains, their resource requirements are inherently limited. As shown in Table {\ref{tab:refe}}, most studies focus on reducing the computation resources required for FL training {\cite{ramanan2020baffle, rathore2019blockdeepnet}} or model verification {\cite{weng2019deepchain}} processes, which are more efficient in private blockchain-empowered FL studies. Nevertheless, private blockchains relying heavily on authentication increase the risk of power concentration and vulnerability to single-point failures. Using private blockchains in closed ecosystems limits the scalability of the embedded edge networks.} 
% Nevertheless, heavily authentication-relied private blockchains exacerbate the issue of power concentration within a single organization, increasing network vulnerability to single-point attacks. Additionally, the requirements for closed ecological application scenarios limit the scalability of blockchain-embedded edge networks.

\subsubsection{Public Blockchain-Empowered FL}
Public blockchains allow everyone to participate in transactions and engage in the consensus process, offering higher scalability over private blockchains. {They are particularly well-suited to accommodate an increasing number of participants with evolving communication technologies}. Kim \textit{et al.} \cite{kim2018device} {propose BlockFL for P2P FL that uses public blockchain to store and verify exchanged model updates, providing defense against attacks while ensuring low latency and high reliability required by future wireless systems}.
% This decentralized architecture trains models in Peer-to-Peer (P2P) without any central organization. 
% Ma \textit{et al.} \cite{ma2022federated} proposed a security learning paradigm with a public blockchain, namely BLADE-FL, to safeguard the learning process against poisoning attacks by malicious users for FL robustness. 
% However, significant computation resource consumption in the mining process and the extensive on-chain model storage severely limit the applicability of public blockchains to MENs. 
A distributed public blockchain-empowered system, namely Biscotti \cite{shayan2020biscotti}, designs a proof of federation to validate models according to peers' contributions. 
% It is challenging for a block with limited storage capability to store larger models while maintaining its format. BAFFLE leveraged a chunk-and-serialized method to ensure the entire model can be stored on blockchain \cite{ramanan2020baffle}. Additionally, 
Rehman \textit{et al.} \cite{ur2020towards} and Zhang \textit{et al.} \cite{zhang2020blockchain} store InterPlanetary File System (IPFS) addresses and clients' registry information instead of the entire model parameters or gradients to reduce the storage resources of public blockchains, respectively. 
% Although the storage required by a single block is reduced, continuous model transactions significantly increase storage demands, posing a major challenge for managing public blockchain on edge devices. 

{Given the global scalability, optimizing computation and storage resources required in public blockchains is more crucial than other structures. As analyzed in Table {\ref{tab:refe}}, a common strategy emerging from representative studies is designing lightweight consensus mechanisms to replace computationally intensive Proof-of-Work (PoW) {\cite{shayan2020biscotti,li2024ppbfl}}. Optimizing the FL training process is another practical approach to reduce the burden of computation resources {\cite{feng2021bafl}}. In terms of storage resources, storing an encrypted identifier instead of the raw model enhances security and efficiency simultaneously {\cite{qu2020decentralized,ur2020towards,li2024ppbfl}}. Herein, IPFS stands out as a promising decentralized storage technique. Nevertheless, continuous model storage still poses a major challenge for managing public blockchain on edge devices.}

\subsubsection{Consortium Blockchain-Empowered FL}
Considering the scalability and efficiency required by edge networks, consortium blockchains, governed collectively by multiple organizations, offer a balance between access control and decentralized management. Guo \textit{et al.} \cite{guo2021lightfed} propose blockchain-empowered LightFed, where the edge nodes upload partial models based on the proposed model splitting and splicing and selective parameter transmission for aggregation and lightweight communication. 
{The flexibility of LightFed makes it highly applicable to edge devices equipped with advanced microprocessors and 5G communication technologies.}
Tang \textit{et al.} \cite{tang2022blockchain} studied how to prevent attacks from malicious nodes and proposed a blockchain-based FL framework {during offloading traffic in the space-air-ground integrated network}. 
% The proposed blockchain shows excellent flexibility in customized device management. 

{Due to the customization of consortium blockchains, the resource consumption of consortium blockchain-empowered FL is always controllable and highly depends on the committee's configuration. In Table {\ref{tab:refe}}, some representative studies focus on optimizing the committee's scale by designing effective elections to reduce communication resource consumption {\cite{yang2022trustworthy, li2020blockchain, chen2020methodology}}. Designing more efficient training and consensus mechanisms can reduce computation resource consumption, as shown in related studies {\cite{chen2020methodology, mothukuri2021fabricfl}}. Li {\emph{et al.}} {\cite{li2020blockchain}} propose a passive storage cleaning method to be activated only when the storage space is exhausted.} However, the complex and overlapping communication topology increases the risks of packet loss and network congestion, potentially affecting performance and security in large-scale edge network deployments.

\subsubsection{Hybrid Blockchain-Empowered FL}
Hybrid blockchains integrate various elements in private, public, and consortium blockchains to offer tailored services for organizations requiring customized accessibility. VFChain \cite{peng2021vfchain} presents a hybrid chain structure for rapid positioning to reduce the query time of edge devices. To reduce the overhead caused by storing entire models, Jin \textit{et al.} \cite{jin2023lightweight} introduce lightweight blockchain-deployed FL for edge networks to validate transactions based on improved proof-of-stake. The scalability and resource efficiency are improved by compressing models by PowerSGD \cite{vogels2019powersgd}. By seeking the balance between privacy control of private blockchain and accessibility of public blockchain, hybrid blockchain is well-suited for scenarios requiring more complex and customized management permissions. PermiDAG based on locally Directed Acyclic Graph (DAG) \cite{lu2020blockchain} uses deep reinforcement learning for node selection to enhance data sharing security and efficiency in the IoVs. 

{Compared to other blockchains, hybrid blockchains have greater customization flexibility to design efficient and secure mechanisms. As analyzed in Table {\ref{tab:refe}}, lightweight blockchains are designed by optimizing computation resource utilization {\cite{chai2020hierarchical, lu2020blockchain}} or reducing storage resource demands {\cite{feng2021blockchain, jin2023lightweight}} by compressing models.} However, a hybrid blockchain with a complex structure presents a challenge of increased maintenance costs in MENs.

Overall, with the increasing network scales, the operational costs of blockchain are a critical factor affecting its deployment and management. Therefore, this paper explores a more scalable and lightweight blockchain-empowered architecture for FL tasks in MENs from three primary required resources: computation, communication, and storage.

\section{Massive Edge Network}
In this section, we detail the elements of MENs and formulate two crucial objectives for latency and {consensus security}. 
\begin{table}[!t]
    \centering
    \renewcommand{\arraystretch}{1.2}
    \caption{Major notations.}
    \label{tab:nota}
    \scalebox{0.9}{
        \begin{tabular}{p{0.36\textwidth}p{0.15\textwidth}}
            \toprule
            \textbf{Description} & \textbf{Notations} \\
            \midrule
            Edge device set & $\mathcal{N}=\{1,\dots,N\}$ \\
            Dataset of device $i$ & $D_i$ \\
            Cluster set & $\mathcal{K} = \{1,\dots, K\}$\\
            Communication rate between devices $i$ and $j$ & $r_{i,j}$\\
            Bandwidth, transmission power & $b$ and $p_i$ \\
            Channel gain between devices $i$ and $j$ & $h_{i,j}$\\
            FL training task & $\Lambda$ \\
            Block & $B$\\
            Computation frequency of device $i$ & $c_i$\\
            Binary variable of device $i$ attributed to cluster $k$ & $\alpha_{i,k}$\\
            Committee member variable of device $i$ in cluster $k$ & $\beta_{i,k}$\\
            \bottomrule
        \end{tabular}
    }
\end{table}
\subsection{Overview of MENs}
\begin{figure*}
    % \centering
    \includegraphics[width = \linewidth]{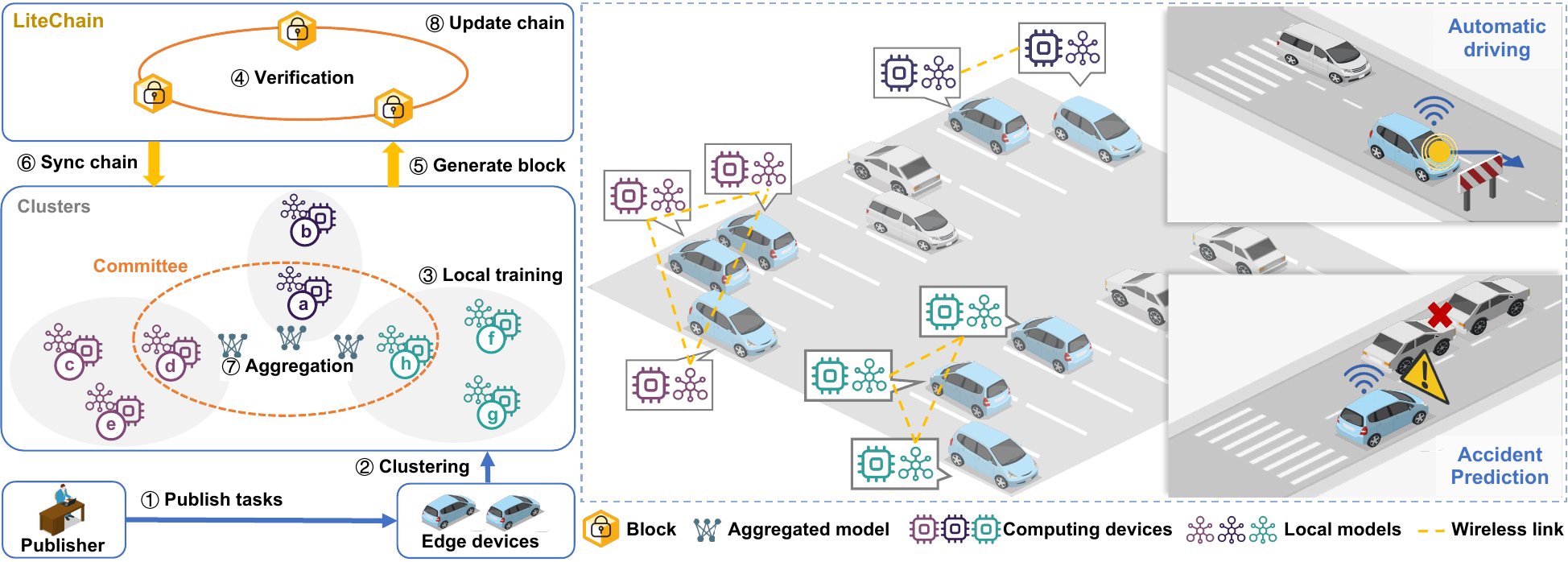}
    \caption{The system overview of LiteChain, including the architecture of LiteChain (left) and LiteChain application in a sample MEN (right). The architecture includes: \ding{192} Task publishers dispatch FL tasks to edge devices; \ding{193} A two-tier LiteChain formation via distributed optimization algorithm; \ding{194} Off-chain verification-based intra-cluster local training; \ding{195} Request on-chain verification via consensus mechanism; \ding{196} Block invocation; \ding{197} LiteChain synchronization; \ding{198} Aggregation of updated models according to LiteChain; \ding{199} After $n$ step updates leading to reputation record updates and redundant storage clearance via commitment.}
    \label{fig:overview}
\end{figure*}

An MEN comprises a large-scale of edge devices with diverse heterogeneous capabilities in computation, communication, and storage. The major notations in this study are elucidated in Table \ref{tab:nota}. Fig. \ref{fig:overview} overviews the system of LiteChain. The left part details the LiteChain framework and the construction process, encompassing the following steps illustrated in the figure.
\begin{enumerate}
    \item An FL training task is distributed by a task publisher to the accessible edge devices. The device set in the MEN is denoted as $\mathcal{N}=\{1,2,\dots,N\}$;
    \item The MEN is reorganized into a hierarchical structure with higher efficiency through the distributed clustering Alg. \ref{alg:cluster} introduced in Section 4.1. The reorganized clusters are denoted by $K\in\mathcal{K}$, where a node is elected in each cluster to construct a committee for on-chain {verification}. Each committee member takes dual roles: a controller of the cluster and a verifier verifying the blockchain;
    \item After $\Phi$ iterations of local training, edge devices forward their models, trained from time $\tau$, to their committee members for off-chain {verification} and synchronous intra-cluster aggregation; 
    \item After intra-cluster aggregation, the committee member requests block upload and initiates a CBFT consensus mechanism for {verification} with the other committee members; 
    \item Upon consensus, the requester records the new block to LiteChain; 
    \item The updated LiteChain will be synchronized to the other devices; 
    \item The committee members aggregate the updated models by a staleness-aware asynchronous inter-cluster aggregation algorithm; 
    \item Following $\chi$ communication rounds, all the committee members clean the stale records, synchronize models' versions, and update the committee members through the proposed update mechanism. 
\end{enumerate}
{The right part of Fig. {\ref{fig:overview}} presents the implementation of LiteChain in a parking lot, leveraging the idle resources of massive vehicles with varying computation capacities for FL tasks. These disparate vehicles are adaptively optimized into multiple self-organizing clusters. Each vehicle trains models based on the image data collected from the vehicular camera during driving. The validity of the trained models is established through the consensus mechanisms. Once training is completed, the collectively trained models are utilized in tasks such as object detection for autonomous driving or accident prediction. Additionally, LiteChain is also adaptable to other massive networks, such as city-wide surveillance for traffic accident detection and drones for urban security monitoring. Furthermore, with the advancement of large model-based tools in distributed networks, LiteChain is also promising in enhancing the security and effectiveness of fine-tuning large models among smart devices. }

\subsection{Latency Metric}
The latency in a blockchain-empowered FL communication round consists of two parts, i.e., training latency and blockchain verification latency. Based on the hierarchical architecture, {the latency of device $i$ in cluster $k$ can be formulated as follows}:
\begin{align}
\label{equ:latency}
    \begin{split}
        T_i =& \underbrace{\sum_{k\in\mathcal{K}}\alpha_{i,k}(\max_{i'\in\mathcal{N}}\{\alpha_{i',k}T_{i',k}^{train}\}+\sum_{j\in\mathcal{N}}\beta_{j,k}(T_{j,k}^{agg}}_{\textit{FL training}}\\
        &+\underbrace{T_{j,k}^{bc}))}_{\textit{verification}}.
    \end{split}
\end{align}
The former term is the intra-cluster latency depending on the maximum device training time $\max_{j\in\mathcal{N}}\{\alpha_{j,k}T_{j,k}^{train}\}$. The latter term $\sum_{j\in\mathcal{N}}\beta_{j,k}(T_{j,k}^{agg}+{T_{j,k}^{bc})}$ is the inter-cluster latency including aggregation latency $T_{j,k}^{agg}$ and blockchain-empowered verification latency $T_{j,k}^{bc}$. $\alpha_{i,k}$ and $\beta_{j,k}$ are binary variables, where $\alpha_{i,k}=1$ indicates device $i$ belonging to cluster $k$.
$\alpha_{i,k}$ needs to satisfy
\begin{equation}\label{C3}
    \alpha_{i,k}\in\{0,1\}, \sum_{k\in\mathcal{K}}\alpha_{i,k} = 1,
\end{equation}
which means that device $i$ can only be clustered into one cluster.
$\beta_{j,k}=1$ indicates that device $j$ is the committee member of cluster $k$, and $\alpha_{j,k}=1$. {That is}
\begin{equation}\label{C4}
    \beta_{j,k} \in\{0,1\}, \sum_{j\in\mathcal{N}}\alpha_{j,k}\beta_{j,k}=1.
\end{equation}
Besides, each cluster can only authorize one device as a committee member, that is,
\begin{equation}\label{C2}
    \left\vert\sum_{j\in\mathcal{N}}\sum_{k\in\mathcal{K}}\beta_{j,k}\right\vert = K.
\end{equation}

\textbf{FL training latency.}
The achievable communication rate between devices $i$ and $j$ can be calculated by Shannon Theory, i.e., 
\begin{equation}
    r_{i,j}=b\log(1+\frac{p_ih_{i,j}}{\sigma^2}),
\end{equation}
where $p_i$ and $h_{i,j}$ denote the transmission power and channel gain, respectively. $\sigma^2$ indicates the noise power. The channel gain $h_{i,j}$ between devices $i$ and $j$ follows the free-space path loss model \cite{huang2020deep}, which can be obtained by:
\begin{equation}
h_{i,j} = A_d(\frac{v^l}{4\pi{f_c}d_{i,j}})^{d_e},
\end{equation}
where $A_d$ indicates the antenna gain, $v^l$ indicates the speed of light, $f_c$ denotes the carrier frequency, and $d_e$ denotes the path loss exponent. Thus, the intra-cluster training latency of device $i$ can be calculated by
\begin{equation}
    T_{i,k}^{train} = \underbrace{\frac{\Lambda^{comp}D_i}{c_i}}_{\textit{local train}}+\underbrace{\frac{\sum_{j\in{K_k}}\beta_{j,k}\Lambda^{size}}{r_{i,j}}}_{\textit{send to committee member}},
\end{equation}
where $\Lambda^{comp}$ indicates the number of floating-point operations required to train on a single data instance depending on the specific model used. 
Note that only one device can be authorized as the committee member, as illustrated in equation (\ref{C4}).
For committee member $j$ in cluster $k$ ($\beta_{j,k}=1$), the aggregation operation includes synchronous aggregation with the cluster members and staleness-aware asynchronous aggregation with the models of the other clusters, which can be denoted as ${\sum_{j\in{K_k}}\beta_{j,k}\Lambda^{agg}}/{c_j}$, where $\Lambda^{add}$ denotes the number of floating-point operations required for aggregation.

\textbf{Verification latency.} The CBFT consensus mechanism is similar to a practical Byzantine fault tolerance mechanism, including preparation, verification, commitment, and reply phases.
% , and shown in Fig. \ref{fig:pbft}. 
The set of committee members can be denoted as $\mathcal{V}=\{i|\beta_{i,k}=1, i\in\mathcal{N},k\in\mathcal{K}\}$. Thus, the verification latency for each phase can be calculated as follows:
% $3*T_j^{broadcast}+T_{j}^{veri}+T_j^{commit}+T_j^{uni}$. 

\begin{itemize}
    \item Preparation phase: Committee member $j$ first broadcasts the block and the model to the other $(K-1)$ committee members for verification. The broadcasting latency can be calculated as  $\theta{(B^{info}+\Lambda^{size})}(K-1)$, where $\theta$ is a pre-defined parameter reflecting the speed of broadcasting and comparison in serial broadcasting \cite{kang2019toward}. $B^{info}$ and $\Lambda^{size}$ denote the size of the block and the model, respectively. 
    \item Verification phase: After receiving the preparation message, the committee members need to verify the received model. The verification latency can be denoted as $\max_{k'\in\mathcal{V}\backslash{j}}\{{\Lambda^{veri}}/{c_{k'}}\}$, where $\Lambda^{veri}$ represents the number of floating-point operations required for verification, and $c_{k'}$ is the computation capacity of committee member $k'$. The verification message will be broadcast to the other committee members, the latency can be calculated by $\theta{B^{info'}}(K-1)$.
    \item Commitment phase: Upon receiving the verification messages, each committee member commits whether it receives at least $\lceil{(2K+1)}/{3}\rceil$ verification messages and then broadcasts a commitment message to the other committee members. The latency for this stage can be represented as: $\theta{B^{info'}}(K-1)+\max_{k'\in\mathcal{V}}\{{B^{com}}/{c_{k'}}\}$. 
    \item Reply phase: Similar to the commitment phase, after receiving at least $\lceil{(2K+1)}/{3}\rceil$ commitment messages, the latency for returning results to the requester is calculated as $\max_{k'\in\mathcal{V}\backslash{j}}\{{B^{info'}}/{r_{k',j}}\}$.
\end{itemize}

A more detailed explanation of the consensus process is provided in Alg. \ref{alg:consensus} in Section 4.3. After verification, the blockchain needs to be synchronized among all the devices. The device set within cluster $k$ can be represented by $\mathcal{N}_k=\{i|\alpha_{i,k}=1, i\in\mathcal{N}\}$.
The synchronization latency can be calculated by
$\max_{i'\in\mathcal{N}_k}\{B^{info'}/{r_{i',j}}\}$.

In summary, the blockchain verification latency can be represented as:
\begin{align}
\begin{split}
    T_{j,k}^{bc}=&\underbrace{\frac{B^{gen}}{c_j}}_{\textit{generate}}+\underbrace{\theta (K-1)(B^{info}+\Lambda^{size}+2B^{info'})}_{\textit{broadcast}}+\\
    &\underbrace{\max_{k'\in\mathcal{V}\backslash{j}}\{\frac{\Lambda^{veri}}{c_{k'}}\} +\max_{k'\in\mathcal{V}}\{\frac{B^{com}}{c_{k'}}\}}_{\textit{verification in Phase 1}\sim\textit{3}}+\underbrace{\max_{k'\in\mathcal{V}\backslash{j}}\{\frac{B^{info'}}{r_{k',j}}\}}_{\textit{unicast result}}.\\
\end{split}
\end{align}

% \subsection{\colorbox{yellow}{Consensus Security Metric}}
\subsection{{Consensus Security Metric}}
Each IoT device is assumed to be a semi-trusted adversary, operating independently without collusion with the other devices. Most participants can obey their duties honestly, while others may be malicious or attacked by some adversaries, leading to incorrect behaviors. The reliability of device $i$ is denoted as $p_i$ ($p_i\in[0,1]$), which can be evaluated by the normalized reputation $r_i$. A detailed explanation of reliability calculation is provided in Section 4.4. {Inspired by {\cite{li2020scalable}}}, the {consensus security} is defined as the probability of achieving a successful consensus that adheres to BFT, as elucidated in Theorem \ref{the:security}.

\begin{theorem}
\label{the:security}
    In a system of $n$ devices with reliability $P=\{p_1, p_2, ..., p_n\}$ of passing the consensus, the success probability with $m$ malicious nodes can be obtained by 
    \begin{equation}
        S_m= \sum_{m'=1}^{\binom{K}{m}}\prod_{j\in\mathcal{V}_{m'}}(1-p_j)\prod_{j\in\mathcal{V}_{m'}^c}p_j,
    \end{equation}
    where $\mathcal{V}_{m'}$ indicates the set of malicious nodes, and $\mathcal{V}_{m'}^c$ denotes the complement of $\mathcal{V}_{m'}$, i.e., the set of normal nodes.
    According to the BFT limit, the {consensus security} can be guaranteed with no more than $1/3$ malicious nodes. {According to the cumulative distribution function {\cite{feller1991introduction}}}, the {consensus security} can be obtained by 
    \begin{equation}
    \label{equ:security}
        S = \sum_{m=0}^{\lfloor\frac{K-1}{3}\rfloor}\sum_{m'=1}^{\binom{K}{m}}\prod_{j\in\mathcal{M}_{m'}}(1-p_j)\prod_{j\in\mathcal{M}_{m'}^c}p_j.
    \end{equation}
\end{theorem}

The proof can be found in Appendix A. The number of clusters for BFT-based {consensus security} needs to satisfy the following constraint:
\begin{equation}\label{C1}
    4\le K\le N.
\end{equation}
The computation complexity is $\mathcal{O}(\sum_{m=0}^{\lfloor\frac{K-1}{3}\rfloor}\binom{K}{m}K)$. When $K=100$, the computation times are more than $5.54\times 10^{28}$, leading to a disaster in large-scale MENs. Although existing approximation methods, such as the Poisson and normal approximations, can estimate results roughly. In this study, we use a calculation formula based on the discrete Fourier transformation (DFT-CF) of distributional eigenfunctions to calculate the {consensus security} proposed in \cite{hong2013computing} in an exact way. According to equation (7) of \cite{hong2013computing}, equation (\ref{equ:security}) can be calculated by:
\begin{align}\label{equ:resecurity}
    \begin{split}
        S =& \frac{1}{K+1}\sum_{k=0}^{K}\sum_{M=0}^{\lfloor\frac{K-1}{3}\rfloor} \exp({-\mathbf{i}\omega kM})\prod_{j = 1}^K\left[1-p_j+\right.\\
        &\left.p_j\exp({\mathbf{i}\omega k})\right],
    \end{split}
\end{align}
where $\omega = {2\pi}/({K+1})$. The proposed algorithms can be found in Alg. A of \cite{hong2013computing}, which takes fast Fourier transformation only once.

\section{LiteChain Design}
 In this section, we provide a detailed explanation of the whole workflow of LiteChain. The workflow of LiteChain is illustrated in Fig. \ref{fig:workflow}. It consists of network initialization (steps 1$\sim$4), intra-cluster training with off-chain {verification} (steps 5$\sim$7), inter-cluster training with on-chain {verification} (steps 8$\sim$14), and secure update consensus (steps 15$\sim$18).

\subsection{Network Initialization}
\begin{figure}[!ht]
    \centering
    \includegraphics[width = 0.9\linewidth]{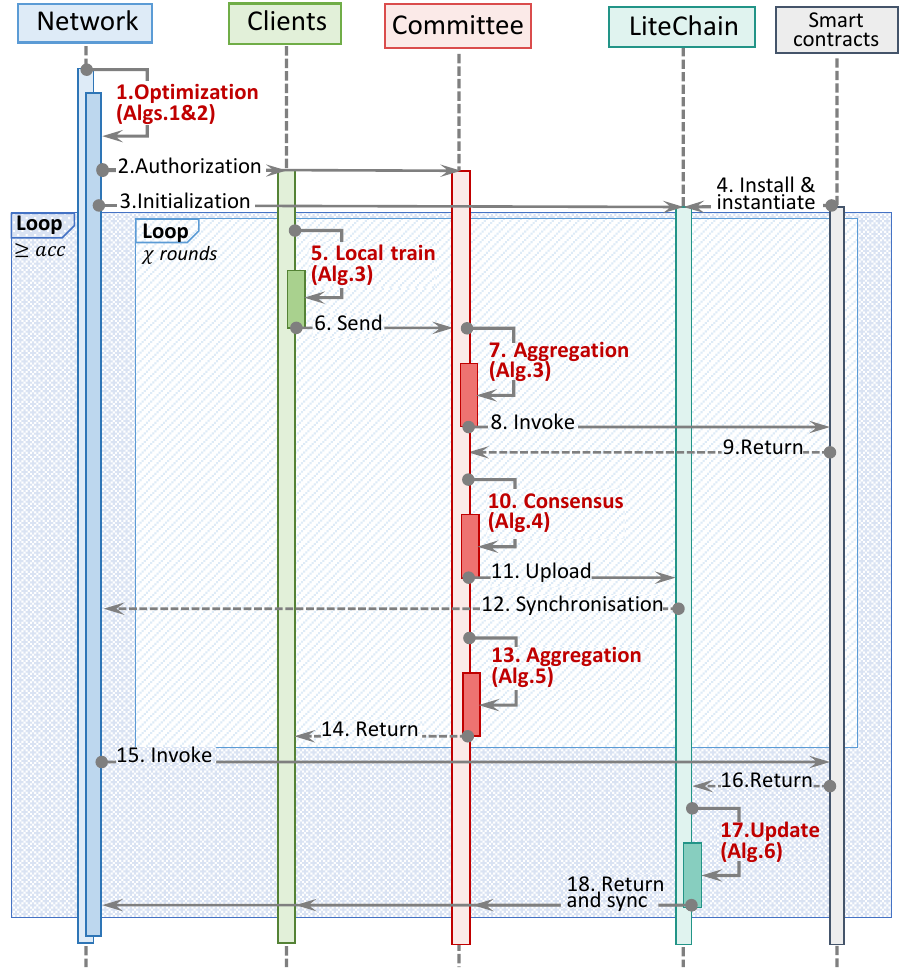}
    \caption{The workflow of LiteChain.}
    \label{fig:workflow}
\end{figure}

Network initialization corresponds to steps 1$\sim$4 in Fig. \ref{fig:workflow}. Before executing the FL training task, the MEN is initially reorganized to a hierarchical network structure for FL training. The network structure is optimized to maximize {consensus security} and minimize latency simultaneously. The bi-objective optimization problem can be formulated as follows:
\begin{equation}
\label{equ:objective}
	\begin{split}
	P: &\max_{K, \alpha_{i,k}, \beta_{i,k}}S\\
       &\min_{K,\alpha_{i,k},\beta_{i,k}}\mathbb{E}[T_{i}].\\
	\end{split}
\end{equation}
subject to the constraints (\ref{C3}), (\ref{C4}), (\ref{C2}), and (\ref{C1}).

The first optimization objective ensures on-chain {consensus security}, calculated by equation (\ref{equ:resecurity}). The second objective intends to minimize the average latency of each device in a single communication round. The latency can be obtained by equation (\ref{equ:latency}). As the optimization problem $P$ shows, the decision variables $K$, $\alpha_{i,k}$, and $\beta_{i,k}$ are coupled deeply. Hence, $P$ is a mixed-integer nonlinear programming problem, known as an NP-hard problem, that cannot be solved in polynomial time.

The bi-objective optimization problem $P$ aims to build a hierarchical network by forming clusters among the edge devices. For highly distributed MENs without a central manager, the limited observable state information restricts the coordination efficiency and optimization precision in self-organizing. We propose a coalition game-based distributed clustering method inspired by \cite{massin2017coalition} to solve the bi-objective optimization problem. The devices are the players in this game. There are two strategies for the devices: \textit{switch} and \textit{remain}. For each step, devices may choose to switch to another cluster or remain in the current cluster. 
The constructed coalition game is detailed as follows:
\begin{enumerate}
    \item Solution: The partitions set in time $t$ presented by $\mathcal{K}^t:=\{{K}_1^t, \cdots, {K}_K^t\}$, where ${K}_k^t$ is the subset of $\mathcal{N}$ and $\cup_{k=1}^{K}{K}_k^t=\mathcal{N}$.
    \item Utility: The utility of cluster $K_k^t$ in time $t$ is defined as $u({K}_k^t)={S^t}/{T_k^t}$ to minimize latency and maximize {consensus security} simultaneously.
    \item Cost: Cost $c(K_k^t)$ presents the punishments for constraints. If constraints are satisfied, $c({K}_k^t)$ is $0$, otherwise, $c({K}_k^t)$ is a large enough value.
    \item Value: Based on utility and cost, the value of cluster $v(K_k^t)$ is defined as $v({K}_k^t)=u({K}_k^t)-c({K}_k^t)$.
    \item Marginal contribution: Value difference of device $i$ to cluster $K_k$ in time $t$ is defined as $i$'s marginal contribution to $K_k$, i.e., $r_i(K_k^t)=v(K_k^t)-v(K_k^t\backslash{i})$. 
    \item Switch operation: The operation $\sigma_{k,l}^t(i)$ denotes switching device $i$ from cluster $K_k^t$ to cluster $K_l^t$;
    \item Switch operation gain: The marginal contribution gap of switching device $i$ between two clusters, denoted as ${G}(\sigma_{k,l}^t(i))=r _i({K}^t_l\cup{i})-r_i({K}_k^t)$.
    \item Preference relation: $\succ$ represents the dominant relationship based on the switch operation gain. If $G(\sigma_1)>G(\sigma_2)$, a switch operation $\sigma_1$ is preferred over $\sigma_2$. $\succ$ represents the dominant relationship based on the switch operation gain, i.e., $\sigma_1\succ\sigma_2$.
    \item Nash-stable: A state is Nash-stable if no switch operation provides a positive gain, that is, $G(\sigma(*)\le0)$, for all operation $\sigma(*)$. The stable state means no single switch can improve the overall system.
\end{enumerate}

\begin{algorithm}
\label{alg:cluster}

    \caption{Pseudo-code of distributed network optimization for LiteChain}
    \KwIn{Device set $\mathcal{N}$}
    \KwOut{Number of cluster $K$, clustering result $\alpha_{i,k}$, committee member result $\beta_{i,k}$}
    Initial partitions $\mathcal{K}^{(0)}=\{\{K_1\}, \cdots, \{K_N\}\}$ to singletons and $\beta_{i,k}=1$ for all $i\in\mathcal{N}$\;
    Initial visited times $i.visit=0$ and switch preference list $\mathcal{P}_i$\;
    % Initial switch operation set $\mathcal{S}$\;
    % Initial operation set $\mathcal{M}=\emptyset$\;
    \While{time slot $t=0,1,\cdots$}{
        \For{each cluster $\mathcal{K}_k^{t}\in\mathbf{K}^{t}$}{
            % \textbf{Phase 1: check state}\;
            \If{$K_k^{t}$ receive regret operation}{
                Change the state of cluster to \textit{available}\;
            }
            % \textbf{Phase 2: propose switch}\;
            \textit{/* Phase 1: propose switch request */}\\
            Select $i$ with the minimum visited times as the candidate device\;
            Obtain the neighbor cluster set $\mathcal{N}^t(i)$ of $i$\;
            Update preference list according to Alg. \ref{alg:prefer}\;
            % \textbf{Phase 3: receive switch}\;
            \textit{/* Phase 2: receive switch request */}\\
            Obtain received switch operation $\sigma_{k,l}^{t}(i)$ in time slot $t$\;
            \If{$G(\sigma_{k,l}^t(i))>G(\sigma_{l,k}^{t-1}(i))$}{
                Regret switch operation $\sigma_{l,k}^{t-1}(i)$\;
                Obtain operation $\sigma_{k,l}^t(i)$ as candidate operation\;
                Change the state of cluster to \textit{occupied}\;
            }
    }
    \If{switch operations set is not empty}{
        Execute switch operations in time slot $t$\;
    }\Else{
        Terminate algorithm\;
    }
    }
    \textbf{Return} number of cluster $K$, clustering result $\alpha_{i,k}$, committee member result $\beta_{i,k}$.

\end{algorithm}

Based on the above settings, the game is presented in Alg. \ref{alg:cluster}. The partitions are initialized with a single node, where each node serves as the committee member of that cluster. Device $i$ maintains a switch preference list $\mathcal{P}_i$ and visited times $i.visit$. List $\mathcal{P}_i$ is ordered based on the switch operation gain among all communicable devices. Every cluster proposes switch operations (lines 8$\sim$11 of Alg. \ref{alg:cluster}) and checks the received switch requests (lines 12$\sim$17 of Alg. \ref{alg:cluster}) in each time slot. Both devices and clusters have two distinct states: \textit{available} and \textit{occupied}. The \textit{occupied} state signifies that a device is currently engaged in another switch operation and inaccessible for additional switch operations. After checking the cluster state, the device with the minimal visited times is selected to propose its first switch operation choice according to the preference list. The first choice determination process is detailed in Alg. \ref{alg:prefer}. Then, the selected device checks the switch operation gain of all the received switch operations. After proposing and receiving the switch requests, the switch operations will be executed. The algorithm terminates when there is no executable switch operation in time $t$, indicating that there is no switch operation in the network with $G(\sigma(*))>0$.

\begin{algorithm}
\label{alg:prefer}
\caption{Update preference list of devices in cluster}
    \For{each neighbor cluster ${K}_{l}^{t}$ of device $i$}{
                Obtain committee member $\beta_{i',k}^{t}$ with the current partition $\mathcal{K}^{(t)}$ and $\alpha_{i,k}^{t}$\;
                Calculate the switch operation gain $G(\sigma_{k,l}^t(i))$\;
                \If{$G(\sigma_{k,l}^t(i))>0$}{
                    Obtain and order the preference list $\mathcal{P}^t_i=\mathcal{P}^t_i\cup\sigma_{k,l}^t(i)$\;
                }
            }
            \While{$\mathcal{P}^t_i$ is not empty}{
                Select first switch choice $\sigma_{k,l^*}^{t}(i)$ satisfying $\sigma_{k,l^*}^{t}(i)\succeq{\sigma_{k,l}^t(i)}$, $\forall{\sigma_{k,l}^t(i)\in\mathcal{P}^t_i}$\;
                % Here
                \If{$G(\sigma_{k,l^*}^{t}(i))>G(\sigma_{k,l'}^{t-1}(i))$ and the nodes in cluster $\mathcal{K}_{l^*}^{t}$ are available}
                {
                    Propose switch operation $\sigma_{k,l^*}(i)$ to cluster $\mathcal{K}_{l^*}^{t}$\;
                }
            }
\end{algorithm}

After finishing Alg. \ref{alg:cluster}, LiteChain is established based on the reorganized network. The initial system with eligible committee members authorizes each node and allocates key pair $(\textit{sk},\textit{pk})$, including the secret key and public key, to each device for record verification during FL training. 

\subsection{Intra-cluster Training with Off-chain {Verification}}

\begin{algorithm}
    \label{alg:train}
    \caption{Intra-cluster training in LiteChain.}
    \For{time $\tau=0,1,\cdots$}{
        \textit{/* Device $i$ in cluster $k$: */}\\
        % \textbf{\textit{LocalTrain($w_{k;\tau}$)}}\\
        \For{training round $\phi=0,\cdots,\Phi$}
        {
          % Compute $\nabla f_i(w_{i,k;\tau,\phi-1})$\;
          $w_{i,k;\tau,\phi}=w_{i,k;\tau,\phi-1}-\eta\nabla f_i(w_{i,k;\tau,\phi-1})$\;
        }
        Send trained model $w_{i,k;\tau,\Phi}$ to committee member $j$\;
        \vspace{3mm}
        \textit{/* Committee member $j$ of custer $k$: */}\\
        % \textbf{ \textit{Intra-clusterAggregation($w_{*,k;\tau,\Phi}$)}}\\
        \For{each model $w_{i,k;\tau,\Phi}$}
        {
        Sample partial test dataset $D_{k;\tau}$\;
        Validate whether the accuracy of $w_{i,k;\tau,\Phi}$ in $D_{k;\tau}$ exceeds the threshold $\mathcal{A}$ and the validity of its signature\;
        Record model {verification} result of $w_{i,k;\tau,\Phi}$\;
        }
        Aggregate all {verified} models by $w_{k;\tau} = \frac{\sum_{i\in{K'}}\vert D_i\vert w_{i,k;\tau,\Phi}}{\sum_{i\in{K'}}\vert D_i\vert}$\;
        Calculate hash value $H(w_{k;\tau})$ of $w_{k;\tau}$\;
        Invoke chaincode to generate block with participant records and $H(w_{k;\tau})$\;
        Request for consensus with Alg. \ref{alg:consensus}.
    }
\end{algorithm}
Intra-cluster training based on the initial models is shown in steps $5\sim7$ of Fig. \ref{fig:workflow}. Alg. \ref{alg:train} represents the pseudocode of intra-cluster training. 
For device $i$ in a cluster, $i$ updates the model by calculating the gradient of the loss function, formulated as:
\begin{equation}
    f_{i}(w_{i,k;\tau,\phi}) = \frac{1}{|D_{i}|}\sum_{j=1}^{|D_{i}|}f_i(w_{j,i,k;\tau,\phi}),
\end{equation}
where $\tau$ and $\phi$ indicate the time of receiving the model and the number of the local update step, respectively. 
After $\Phi$ rounds of local training, the trained model $w_{i,k;\tau,\Phi}$ is sent to committee member $j$ of cluster $k$ (lines 3$\sim$6 of Alg. \ref{alg:train}). 

For {consensus security} considerations, committee member $j$ tests the quality of the received models before intra-cluster aggregation. According to specified objectives, diverse quality metrics can be used to measure model quality in LiteChain. {In this study, we use test accuracy to measure model quality based on a small sample test dataset to filter out malicious participants with low-quality models. If the test accuracy exceeds the pre-defined accuracy threshold {$\mathcal{A}=1/L$} (line 10 of Alg. {\ref{alg:train}}), the verified model records will be written in transactions with a participant's signature. 
% The verified model records will be written in transactions with a participant's signature if the test accuracy exceeds the pre-defined accuracy threshold {$\mathcal{A}=1/L$} (line 10 of Alg. {\ref{alg:train}}). This is an initial step to filter out malicious users with low-quality models. 
Herein, {$L$} denotes the number of labels, and {$\mathcal{A}$} represents the accuracy of choosing randomly.} After validating all received models, the {verified} ones will be aggregated by FedAvg \cite{mcmahan2017communication}. Committee member $j$ calculates the hash value of the aggregated model to serve as its unique model identifier. Committee member $j$ invokes a smart contract to generate a block storing the model identifier to verify the received content.

\subsection{Inter-cluster Aggregation with On-chain {Verification}}

\begin{algorithm}
    \label{alg:consensus}
    \caption{Comprehensive Byzantine fault tolerance consensus}
    \KwIn{Block with $H(w_{k;\tau})$, model $w_{k;\tau}$}
    \KwOut{Consensus result}
    % \textit{/*Pre-Prepare*/}\\
    Broadcast block to all committee members\;    
    % \textit{/*Prepare*/}\\
    \For{all committee member $j\in\mathcal{V}$}
    {
    \If{block is {verified} \textbf{and} $H(w_{k;\tau})$ is not duplicated }
    {
        \If{$w_{k;\tau}$ satisfies minimum accuracy threshold $\mathcal{A}$}
        {
            Broadcast verification message to all committee members\;
        }
    }
    \textbf{else} break\;
    }
    % \textit{/*Commit*/}\\
    \For{all committee member $j\in\mathcal{V}$}
    {
        \If{receive $\lceil\frac{2K+1}{3}\rceil$ verification message}{
            Broadcast commit message to other committee members\;
        }
        \textbf{else} break\;
    }
    % \textit{/*Reply*/}\\
    \For{all committee member $j\in\mathcal{V}$}
    {
        \If{receive $\lceil\frac{2K+1}{3}\rceil$ commit message}{
        Uni-cast {verification} result to requester\;
        }
        \textbf{else} break\;
    }
\end{algorithm}
After intra-cluster aggregation with off-chain {verification}, inter-cluster training with on-chain {verification} for recording on LiteChain will be executed corresponding to steps 8$\sim$14 in Fig. \ref{fig:workflow}. Committee member $j$ requests to initiate CBFT to validate and record the block on LiteChain. The pseudocode of CBFT can be found in Alg. \ref{alg:consensus}.
Initially, the block and the aggregated model are broadcast to the other committee members to validate the signature and certificate. Then, committee members query Litechain to check the uniqueness of the received model identifier with the previous blocks to prevent replay attacks (line 3 of Alg. \ref{alg:consensus}). Then, committee member $j$ verifies the model quality with the same method leveraged in off-chain {verification} (line 4 of Alg. \ref{alg:consensus}). If the block and the model are verified, the verification message will be broadcast to other committee members for commitment (line 5 of Alg. \ref{alg:consensus}). Once receiving the $\lceil(2K+1)/3\rceil$ verification message, the commit message will be broadcast to the other committee members (lines 10$\sim$15 of Alg. \ref{alg:consensus}). Once a committee member receives \(\lceil(2K+1)/3\rceil\) commit messages, a consensus success message will be uni-cast to the requester (lines 16$\sim$21 of Alg. \ref{alg:consensus}). The consensus process finishes. 
\begin{algorithm}
    \label{alg:inter}
    \caption{Inter-cluster training in LiteChain}
    \For{time $t=0,1,\cdots$}{
        \textit{/* Committee member $j$ of cluster $k$: */}\\
        % \textbf{\textit{Inter-clusterAggregation($\tau,t,w_{k;\tau,\Phi}$)}}\\
        Check updated models until $t-1$ in LiteChain\;
        Calculate the staleness $s_{k;\tau,t}=s(t-\tau+1)^{-q}$\;
        Aggregate updated models with $w_{k;t} = \sum_{k'\in\mathcal{K}}s_{k';\tau',t'}w_{k';t'}-s_{k;\hat{\tau},\tau}w_{k;\hat{\tau}}+s_{k;\tau,t}w_{k;\tau}$\;
        Assigned updated models to devices within cluster $k$\;
    }
\end{algorithm}

% After the validator $j$ obtains the aggregated model, device $j$ constructs the block with the packed transaction and proposes the verification request to other committee members. The latency of generating a block can be calculated by $B^{gen}/f_j$, where required computation resources $B^{gen}$ includes verify signature and hash calculation operations. 
The successfully consensed blocks will be recorded in LiteChain and synchronized with all the other devices. Committee member $j$ accesses the latest models of the other committee members by querying LiteChain. To maximize utilization of the computation resources and reduce the waiting time, committee member $j$ of cluster $k$ adopts equation (\ref{equ:stale}) for staleness-aware asynchronous aggregation (line 5 of Alg. \ref{alg:inter}).
\begin{equation}
\label{equ:stale}
    w_{k;t} = \sum_{k'\in\mathcal{K}}s_{k';\tau',t'}w_{k';t'}-s_{k;\hat{\tau},\tau}w_{k;\hat{\tau}}+s_{k;\tau,t}w_{k;\tau}.
\end{equation}
Equation (\ref{equ:stale}) represents the updates contributed by the committee member at time $t$, and $s_{k;\tau,t}$ denotes the weight related to the model staleness of cluster $k$ from time $\tau$ to current $t$. According to \cite{xie2019asynchronous}, the staleness weight of cluster $k$ is defined as follows:
\begin{equation}
    s_{k;\tau,t} = s(\tau-t+1)^{-q},
\end{equation}
parameterized by $q>0$. In our work, we set $q={1}/{2}$ \cite{stripelis2022semi}. The updated models are assigned to the devices within cluster $k$ for the following training.

\subsection{Secure Update Consensus}
As FL training is iterating, a static committee raises the risk of power centralization, and increases the potential for internal collusion. Outdated models increase the storage burden and contribute little to future training. Therefore, we propose a secure update consensus mechanism to maintain and update LiteChain periodically. This approach promotes the efficiency and security of training. The secure update process of LiteChain corresponds to steps 15$\sim$18 in Fig. \ref{fig:workflow}.
The details of the secure update consensus procedure are illustrated in Alg. \ref{alg:secure}.

\begin{algorithm}
    \label{alg:secure}
    \caption{LiteChain update consensus mechanism}
    Device $i$ request head $j$ of cluster $k$ to update blockchain\;
    \For{committee member $j\in\mathcal{K}$}{
    Aggregate latest models in $t-\chi$ to $t$\;
    Calculate the reputation of devices within cluster $k$ based on historical participation records\;
    Broadcast model and updated reputation information to other committee members\;
    }
    \For{committee member $j\in\mathcal{K}$}{
        \If{receive $\lceil\frac{2K+1}{3}\rceil$ {verified} update message}{
            Broadcast commit message to other committee members\;
        }
        \textbf{else} break\;
    }
    \For{committee member $j\in\mathcal{K}$}{
        \If{receive $\lceil\frac{2K+1}{3}\rceil$ {verified} commit message}{
            Delete stored staleness models\;
            Sync updated model and continue training\;
            Update the reliability of each device\;
            Re-elect committee members of each cluster according to objectives (\ref{equ:objective})\;
        }
        \textbf{else} break.
    }
\end{algorithm}

After $\chi$ communication rounds or the termination accuracy is achieved, the committee will initiate a consensus mechanism to update LiteChain. The latest update records in LiteChain will be aggregated by each committee member for verification (line 3 of Alg. \ref{alg:secure}). The reputation of all the devices is updated based on the participation history recorded in LiteChain (line 4 of Alg. \ref{alg:secure}). The devices can earn reputation rewards by successfully generating a block or achieving a consensus, represented by $R^{b^+}$ and $R^{b^-}$, respectively. We set $R^{b^+} \gg R^{b^-}$ to prevent a committee member from dominating the reputation rewards through participating in the consensus process. The block reward $R^{b^+}$ obtained by the committee member is allocated based on devices' participation recorded in LiteChain according to its data contribution, i.e.,
\begin{equation}
    R_k(i)=\frac{\vert D_i\vert}{\sum_{i\in\mathcal{K}_k}\vert D_i\vert}R^{b^+}.
\end{equation}

This updated message will be broadcast to the other committee members for commitment (line 5 of Alg. \ref{alg:secure}). Upon receiving $\lceil\frac{2K+1}{3}\rceil$ update messages, a commit message is broadcast to all the committee members (lines 7$\sim$12 of Alg. \ref{alg:secure}). If $\lceil\frac{2K+1}{3}\rceil$ commit messages are {verified}, the committee members prune obsolete data and synchronize with the other devices for the following training (lines 15, 16 of Alg. \ref{alg:secure}). Meanwhile, the committee constitution is refreshed based on the updated reputation records according to the bi-objectives (\ref{equ:objective}) to ensure the blockchain reflects the latest information.
If achieving consensus, a new committee will be randomly constituted to re-execute the consensus Alg. \ref{alg:secure} until a consensus is reached. This iterative approach ensures integrity and resilience of the consensus process.

\section{Theoretical Analysis}
This section delves into the proposed communication complexity and the convergence of the proposed algorithms. Except for efficiency, we analyze their resilience to adversarial threats against model poisoning and single-point-of-failure attacks in MENs.

\subsection{Reduced communication complexity}
Communication complexity indicates the total communication bits that need to be exchanged for certain tasks \cite{yao1979some}. After optimizing the network topology with Alg. \ref{alg:cluster}, the expected reduction of communication complexity is analyzed in Theorem \ref{the:sys_com}.

\begin{theorem}
    \label{the:sys_com}
    % For the whole LiteChain system, regardless of the number of clusters, for a round of FL updates (i.e. all nodes complete a round of model training), the two-layer framework has less communication complexity than the one-layer framework, and the maximum reduced communication complexity is 
    We define one-round training as all nodes finishing one round of local training and sending the model to the aggregator. From the original one-tier network to the two-tier network reorganized by Alg. \ref{alg:cluster}, the maximal expected reduction of the communication complexity in one-round training is $\Lambda^{size}(N^2-\frac{3N}{2})+\bar{B}^{size}(2N^2+N-36)$. $\Lambda^{size}$ and $\bar{B}^{size}$ represent the model size and the expected values of the block size, respectively. When $N\ge4$, the maximal expected reduction will be greater than 0. As $N$ increases, its effectiveness on reducing communication complexity becomes increasingly pronounced. 
\end{theorem}
The proof can be found in Appendix B. Theorem \ref{the:sys_com} proves the theoretical pinnacle of efficiency gains achievable in communications through implementing Alg. \ref{alg:cluster} for network optimization.

\subsection{Convergence Analysis}
This subsection encompasses the convergence analysis for distributed clustering (Alg. \ref{alg:cluster}) and the implemented hierarchical FL algorithms (Algs. \ref{alg:train} and \ref{alg:inter}). 

\subsubsection{Convergence Analysis for Algorithm \ref{alg:cluster}}
% In LiteChain, the network includes intra-cluster and inter-cluster parts after optimization using Algorithm \ref{alg:cluster}. 
In Alg. \ref{alg:cluster}, we define the Nash-stable state as no switch operation providing positive switch gain. When the algorithm obtains the unique Nash-stable solution, it can be considered to have converged. The uniqueness of the Nash-stable solution is proved in Theorem \ref{theo:con}.
% As the defined Nash-stable situation, we prove the convergence of the proposed distributed clustering algorithm as the Theorem \ref{theo:con} shown.

\begin{theorem}
\label{theo:con}
For the initial singleton partition $\mathcal{K}^{(0)}$, Alg. \ref{alg:cluster} maps to a sequence of switch operations which converges in a finite number of iterations to a unique Nash-stable final partition $\mathcal{K}^{(*)}$. 
\end{theorem}
The proof can be found in Appendix C. 

\subsubsection{Convergence Analysis of Hierarchical FL Training}
As shown in Algs. \ref{alg:train} and \ref{alg:inter}, the hierarchical FL integrates intra-cluster training and inter-cluster training. Intra-cluster training leverages synchronous FL aggregation within a cluster, while the inter-cluster employs a peer-to-peer staleness-aware asynchronous aggregation similar to the settings in \cite{stripelis2022semi}.
We follow the common assumptions \cite{xie2019asynchronous} on the loss function to analyze the convergence of our implemented hierarchical training. The first assumptions are listed as follows: 
\begin{assumption}
\label{ass1}
    It is assumed that the loss function $f$ satisfies the following conditions:
    \begin{enumerate}
        \item $f$ is $\mu$-weakly convex, where $\mu>0$, $g(w)=f(w)+\frac{\mu}{2}\Vert x\Vert^2$ is convex;
        \item $f$ is $\mathcal{L}$-smoothly, where $\mathcal{L}>0$;
        \item $f$ exists at least one solution $x^*$ for global optimization that can minimize the loss function.
    \end{enumerate}
\end{assumption}
Considering that the updates of gradients and parameters will not grow indefinitely, we assume that both gradients and weights are bounded, as shown in Assumption \ref{ass2} \cite{xie2019asynchronous}.
\begin{assumption}
    \label{ass2}
    We assume $\Vert\nabla{f}(w)\Vert^2\le{Q_1}$ and $\Vert\nabla{f}(w_{t,\phi;k,i,j})\Vert^2\le{Q_2}$ for $\forall{w}\in{\mathbb{R}^d}$. The weights are bounded by $\Vert w\Vert^2\in\left[0, W\right]$.
    % \begin{align}
    %     \begin{split}
    %         \Vert\nabla{f}(w)\Vert^2&\le{Q_1}, \forall{w}\in{\mathbb{R}^d};\\
    %     \Vert\nabla{f}(w_{t,\phi;k,i,j})\Vert^2&\le{Q_2}, \forall{w}\in{\mathbb{R}^d};\\
    %     \Vert w\Vert^2&\in\left[0,W\right]
    %     \end{split}
    % \end{align}
\end{assumption}

% We assume that the staleness is bounded by Assumption \ref{ass3}.
% \begin{assumption}
% \label{ass3}
%     The staleness $t-\tau$ at the validator is bounded:
%     \begin{equation}
%         t-\tau\le{\mathcal{T}}.
%     \end{equation}
% \end{assumption}
First, we bound the loss function gap over local training under Assumption \ref{ass1} in Lemma \ref{le:lu} by telescoping one iteration update to $\Phi$ iterations.
\begin{lemma}
\label{le:lu}
    (Local Update) Each device update $\Phi$ iterations, the loss function satisfies:
    \begin{align}
        \begin{split}
            \mathbb{E}\left[F(w_{\tau,\Phi})\right]\le&\mathbb{E}\left[F(w_{\tau,0})\right]-\eta\sum_{\phi=1}^{\Phi}\Vert\nabla{F}(w_{\tau,\phi-1})\Vert^2\\
            &+\frac{\mathcal{L}\eta^2Q_1\Phi}{2}.
        \end{split}
    \end{align}
\end{lemma}
The proof can be found in Appendix D.
After local training, the committee members aggregate the intra-cluster models with FedAvg and then aggregate with the models from the other clusters. We deduce the recursion for the staleness in Lemma \ref{le:st}.
\begin{lemma}\label{le:st}    
(Staleness Recursion) $w_{\tau,\Phi;k}$ is the updated model of committee member $k$ based on the received model in time $t-1$. At time $t$, committee member $j$ trains the model received at time $\tau$. The term $t-\tau$ represents the staleness of the current model. We assume the staleness is bounded by $t-\tau\le{\mathcal{T}}$. The upper bound of the updated model satisfies: 
\begin{align}
    \mathbb{E}\left[\Vert{w}_{\tau} - w_{t-1}\Vert^2\right]\le&{\mathcal{T}^2}s^2\Phi^2\eta^2Q_1;\\
    \mathbb{E}\left[\left\Vert{w}_{\tau} - w_{t-1}\right\Vert\right]\le&\mathcal{T}s\Phi\eta\sqrt{Q_1}.
\end{align}
\end{lemma}
The proof can be found in Appendix E. 

Under Assumptions \ref{ass1}, \ref{ass2} and Lemmas \ref{le:lu}, \ref{le:st}, the upper bound of loss function gradient over time can be further deduced as follows:
\begin{theorem}\label{the:convergence}
    For $L$-smooth and $\mu$-weakly convex loss function $F$, after running training algorithm, we obtain:
    \begin{align}
        \begin{split}
            &\min_{t=0\cdots T-1}\mathbb{E}\left[\|\nabla F(w_{t})\|^2\right]\\
            \leq &\frac{\mathbb{E}\left[F(w_{0}) - F(w_{T})\right]}{s \eta T \Phi} +\frac{\mathcal{L}\eta Q_1}{2}+\mathcal{T}s\sqrt{Q_1Q_2}\\
            &+\frac{\mathcal{L}\mathcal{T}^2s^2\Phi\eta Q_1}{2}+\frac{\mu W}{s\eta\Phi}.
        \end{split}
    \end{align}
    % When condition $\mathcal{T}^2s^3Q_2\Phi\ge2\mu W\mathcal{L}(1+\mathcal{T}^2s^2\Phi)$ is satisfied, model can converge by adjusting the learning rate $\eta$.
    The constant terms can be formulated as a quadratic function in terms of $\eta$. For this quadratic equation, when the discriminant $\Delta\ge0$ is satisfied, i.e., $\mathcal{T}^2s^3Q_2\Phi\ge2\mu W\mathcal{L}(1+\mathcal{T}^2s^2\Phi)$, there exists learning rate $\eta$ letting the minimum expected squared norm of the gradients asymptotically approaches 0.
\end{theorem}
The proof can be found in the Appendix F. The proof is extended from the convergence proof in \cite{xie2019asynchronous} and \cite{tan2022fedproto}.

\subsection{Security Analysis}
LiteChain framework employs a P2P training mode, where its participants retain absolute autonomy over their private data. The private data would not be shared with any third parties. This section analyzes the robustness of LiteChain under two attacks: model poisoning and single-point-of-failure attacks.

\subsubsection{Model Poisoning Attacks}
In LiteChain, model poisoning attacks may occur during intra-cluster and inter-cluster communication. For intra-cluster communication, model quality evaluation before transaction engagement is a frontline of defense against model poisoning attacks. Considering the generality of LiteChain, any model quality measurement can be implemented in LiteChain according to specific training objectives.

For inter-cluster communication, the unique hash-based model identifier can verify the correctness and integrity of the received models. 
% In MENs, it is uncommon for two devices to have training datasets with the same distribution and end up with identical models after training. 
By utilizing distributed clustering, no two clusters are identical in LiteChain. Consequently, even if the participants are identical, the models proposed by two committee members after intra-cluster training Alg. \ref{alg:train} will vary. By implementing the existing SHA-256 and SHA-512 technologies, the uniquely computed hash value of the proposed models serves as a distinctive model identifier to validate the correctness and integrity of the received models by comparing the model identifiers stored in the blockchain blocks. From a long-term perspective, rational participants are motivated to behave honestly to avoid poor reputation to get models.

% \subsubsection{Replay Attacks}
% Incentivized by obtaining more rewards, malicious participants could send stale models to aggregators without training. In our LiteChain, the unique model identifier is employed to detect replayed content, a critical feature for mitigating the risk posed by malicious participants. From a long-term perspective, participants are motivated to choose honest behavior to avoid a reputation normalized below $R^{down}$, resulting in the unavailability of mods.

\subsubsection{Single Point of Failure}
Prior studies on consortium blockchain for FL depend on a trusted committee, raising the risks of single-point-of-failure and power centralization. To address these issues, we propose Alg. \ref{alg:cluster} to elect committee members with high reliability as the committee. The committee will be re-elected periodically to ensure maximal {consensus security} and prevent excessive power accumulation in a few devices. 
The method for calculating reputation is transparent, allowing any device to verify the correctness of the received feedback using the recorded data. If the feedback is disputed, any device can initiate arbitration to freeze the reputation temporarily. The system utilizes verifiable random functions to select a random organization to arbitrate feedback according to the historical records in LiteChain.
Owing to the periodic updates of the committee and high transparency of transactions, attackers cannot disrupt the overall system functionality by targeting at a single or a few nodes.

\section{Performance Evaluation}
This section presents the performance evaluation of LiteChain. First, we detail the experimental settings. Then, we assess the performance from the perspectives of cost and security.

\subsection{Experiment Settings}
LiteChain was built with Hyperledger Fabric 1.4.6 and Python 3.8. We utilized PyTorch 2.0.1 to train FL models and Go 1.13.8 to define smart contracts for model transactions. We implemented \textit{fabric-sdk-py} 1.0 \cite{fabric} for blockchain operations by Python. LiteChain was deployed on a server with four NVIDIA GeForce RTX 3090 GPUs, an AMD EPYC 7313P 16-Core CPU running at 1500MHz, and 251GB of RAM. The coordinates of massive edge devices were randomly sampled in a $1 km\times 1 km$ square. We utilized Docker 24.0.1 to generate containers to simulate independent nodes in one server. In order to simulate different computation capabilities, the experiment was set up in four cases: using CPU, 1 GPU, 2 GPUs, and 4 GPUs to train the models. We employed serial training to simulate multiple devices and enable the switch of computation capabilities among different devices during training. While some nodes stored the model on the CPU and performed local training, others copied the model to the GPU in their turns. In particular, for those using multiple GPUs, they first replicated the model to the GPUs, then scattered the data in an even split, and eventually gathered the outputs across the GPUs. To minimize measurement errors in device execution time, the devices' training latency was averaged over 100 test runs for each computation capability case. 

% The employed ResNet9 architecture comprises four primary convolution blocks, each integrating a convolution layer, batch normalization, and a ReLU layer, with some including additional max pooling. It also features two residual blocks, each encompassing two convolution blocks. The architecture is completed with a classifier block that consists of max pooling, flattening, and a linear layer.

All the experiments are demonstrated with the CIFAR10 dataset \cite{krizhevsky2009learning}. We implement a ResNet9 model to do classification tasks. The model has four main convolution blocks (each with a convolution, batch normalization, and ReLU layer, except some with an additional max pooling layer), two residual blocks (each containing two convolution blocks), and a classifier block with max pooling, flattening, and a linear layer. Experiments are conducted on IID datasets splitted by Dirichlet distribution with $\alpha = 5$ and non-IID datasets with $\alpha=0.2$. The settings of communication mainly refer to \cite{huang2020deep}. The parameters are tabulated in TABLE \ref{tab:para}.

\begin{table}[!t]
    \centering
    \renewcommand{\arraystretch}{1.2}
    \caption{Default parameters \cite{kang2019toward,huang2020deep}.}
    \label{tab:para}
    \scalebox{1}{ \begin{tabular}{p{6cm}p{1.5cm}}
    \toprule
        \textbf{Parameter} & \textbf{Value} \\
    \midrule
        Broadcast timeout $MaxBroadcast$ & 300 seconds \\
        Broadcast coefficient $\theta$ & 0.5 \\
        Antenna gain $A_d$  & 4.11 \\
        Carrier frequency $f_c$ & 915 MHz \\
        Pathloss exponent $d_e$& 2.8 \\
        Speed of light $v^l$ & 3$\times$$10^8$ m/s \\
        Batch size of device's model training & 128 \\
        Epochs of device's model training & 1 \\
        Learning rate $\eta$ & 0.001 \\
    \bottomrule
    \end{tabular}}
\end{table}

In our study, we compare LiteChain with one-tier (without network optimization) blockchain-empowered FL (FLC) and lightweight blockchain-empowered secure and efficient federated learning (BEFL) \cite{jin2023lightweight}. All comparative blockchain-empowered frameworks were implemented in Hyperledger Fabric, with the same model structure, to eliminate discrepancies stemming from different blockchain platforms. The construction details of the comparative blockchain-empowered FL are outlined as follows:
\begin{itemize}
    \item FLC-model: The one-tier blockchain-empowered FL with FedAvg that stores the whole model for {verification}, which is a common setting in the existing studies \cite{tang2022blockchain,peng2021vfchain}.
    \item FLC-hash: The one-tier blockchain-empowered FL with FedAvg that stores the hash value of the models as an identifier which is the same as LiteChain.
    \item BEFL \cite{jin2023lightweight}: Lightweight blockchain-empowered secure and efficient FL system with the powerSGD algorithm for model compression \cite{vogels2019powersgd}. The committee size is 15, the minimum default size in their settings.
\end{itemize}

\subsection{Cost Evaluation}
This section evaluates the latency and storage cost of LiteChain under different scales of MENs. We assess the scalability and cost-efficiency across 50 to 300 devices in IID and non-IID datasets.

\subsubsection{Latency Evaluation}
\begin{figure*}[ht]
    \centering
    \subfigure[50 devices]
    {
        \includegraphics[scale=0.19]{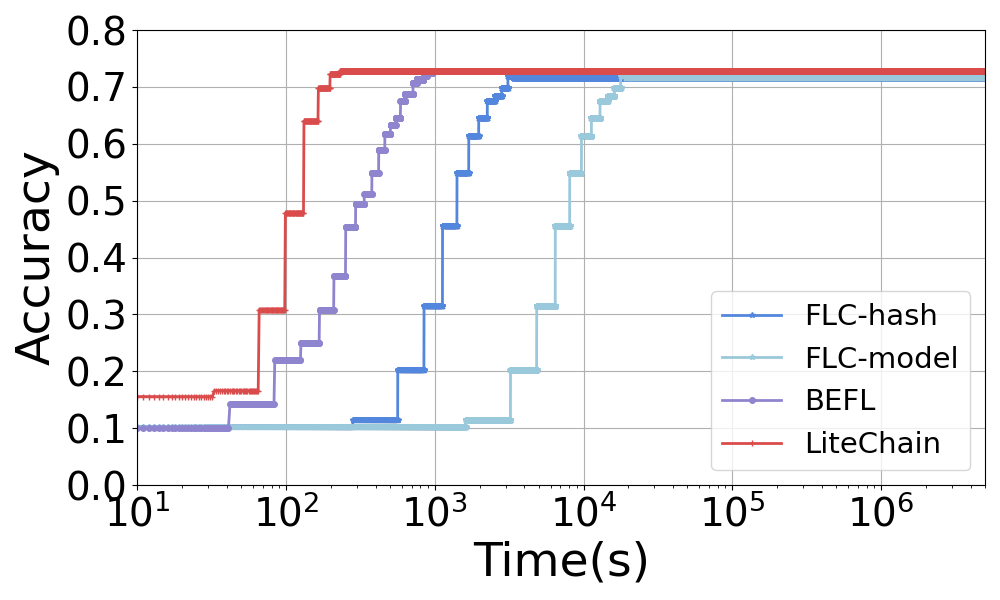}
        \label{s50_5}
    }
    \subfigure[100 devices]
    {
        \includegraphics[scale=0.19]{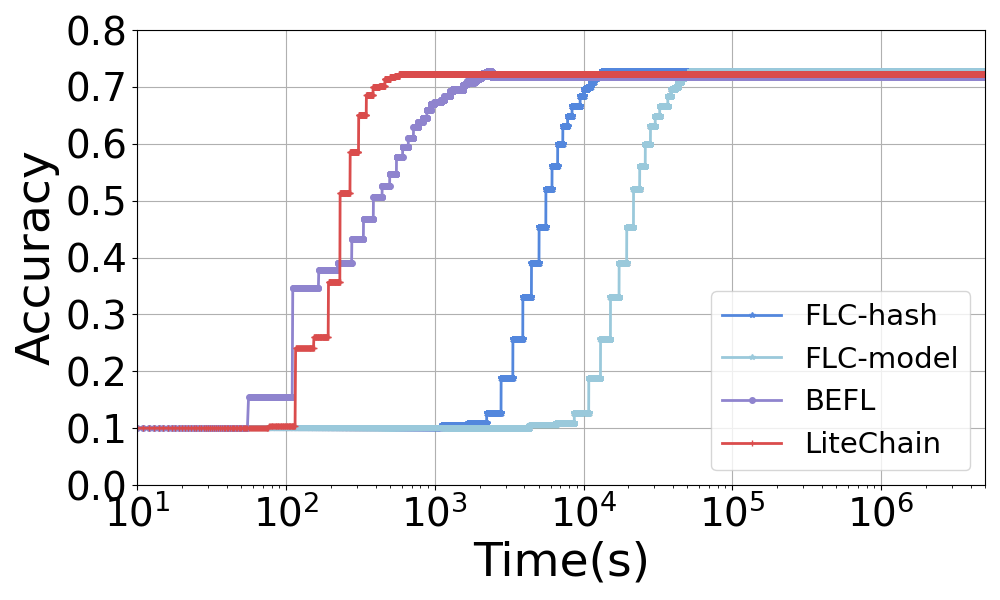}
        \label{s100_5}
    }
    \subfigure[150 devices]
    {
        \includegraphics[scale=0.19]{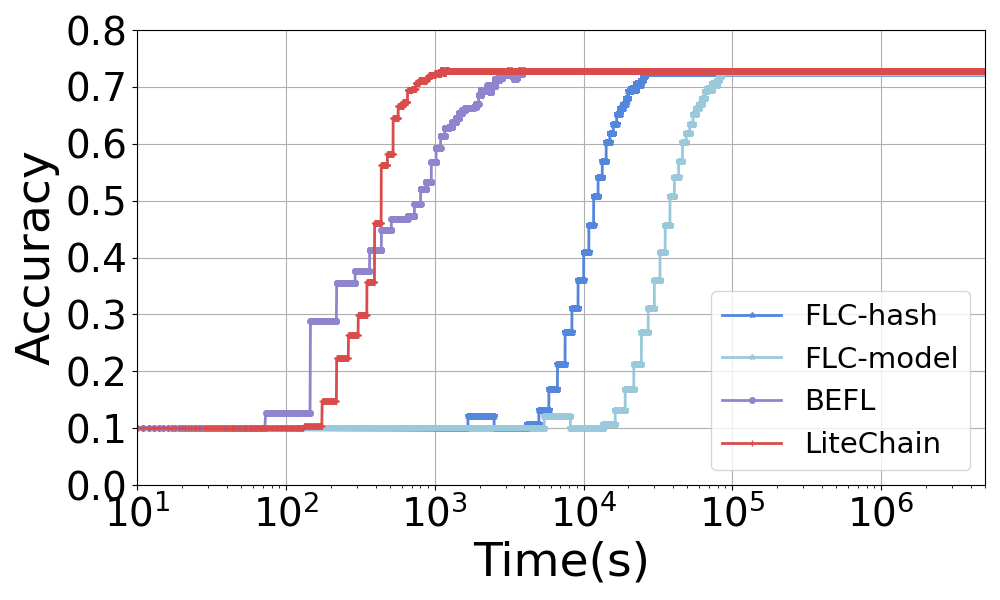}
        \label{s150_5}
    }
    \subfigure[200 devices]
    {
        \includegraphics[scale=0.19]{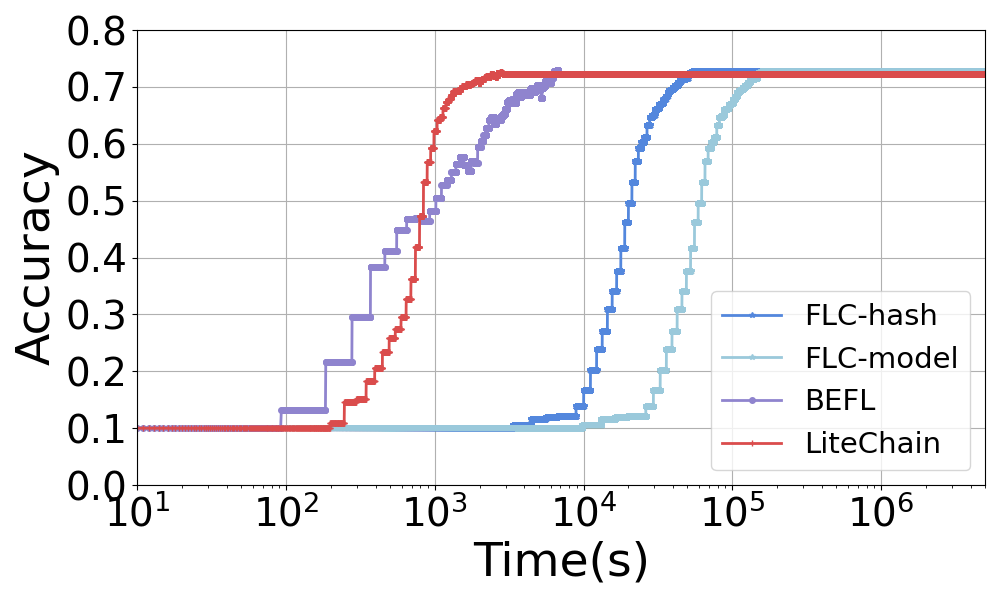}
        \label{s200_5}
    }
    \subfigure[250 devices]
    {
        \includegraphics[scale=0.19]{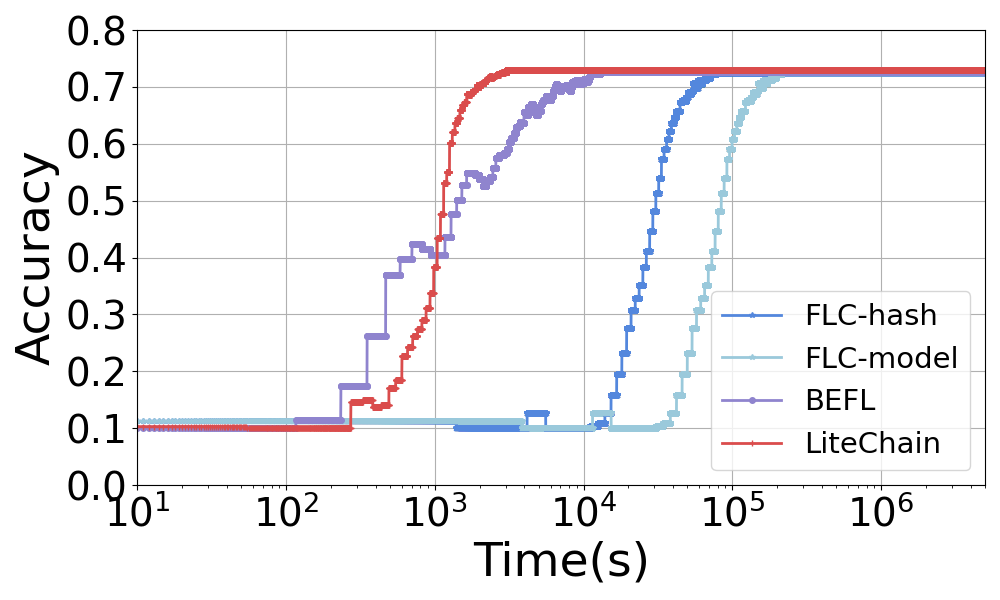}
        \label{s250_5}
    }
    \subfigure[300 devices]
    {
        \includegraphics[scale=0.19]{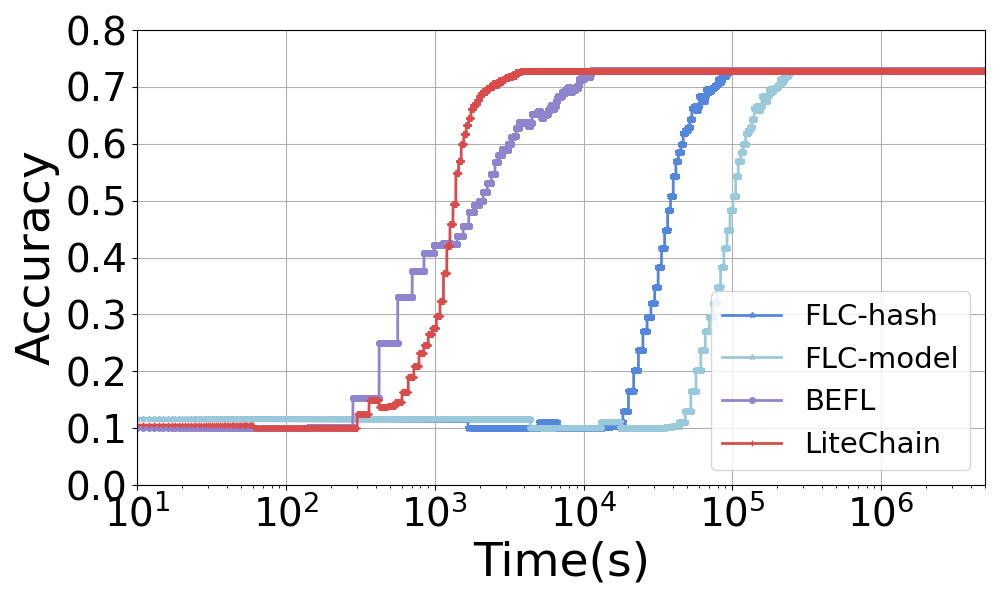}
        \label{s300_5}
    }
    \caption{Accuracy over time (in seconds) with IID dataset.}
    \label{fig:AccTime5}
\end{figure*}

\begin{figure*}[ht]
    \centering
    \subfigure[50 devices]
    {
        \includegraphics[scale=0.19]{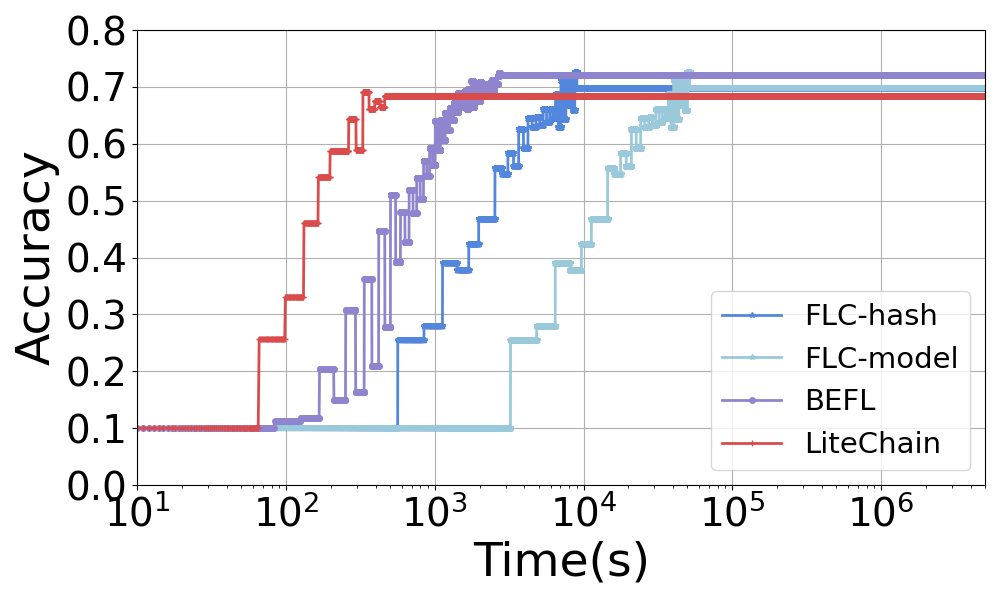}
        \label{s50_02}
    }
    \subfigure[100 devices]
    {
        \includegraphics[scale=0.19]{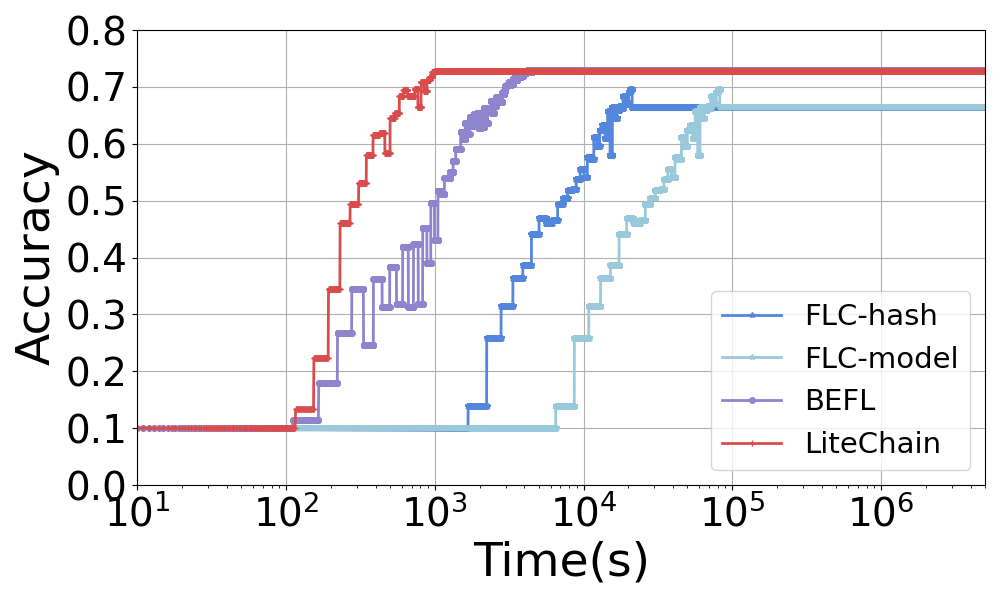}
        \label{s100_02}
    }
    \subfigure[150 devices]
    {
        \includegraphics[scale=0.19]{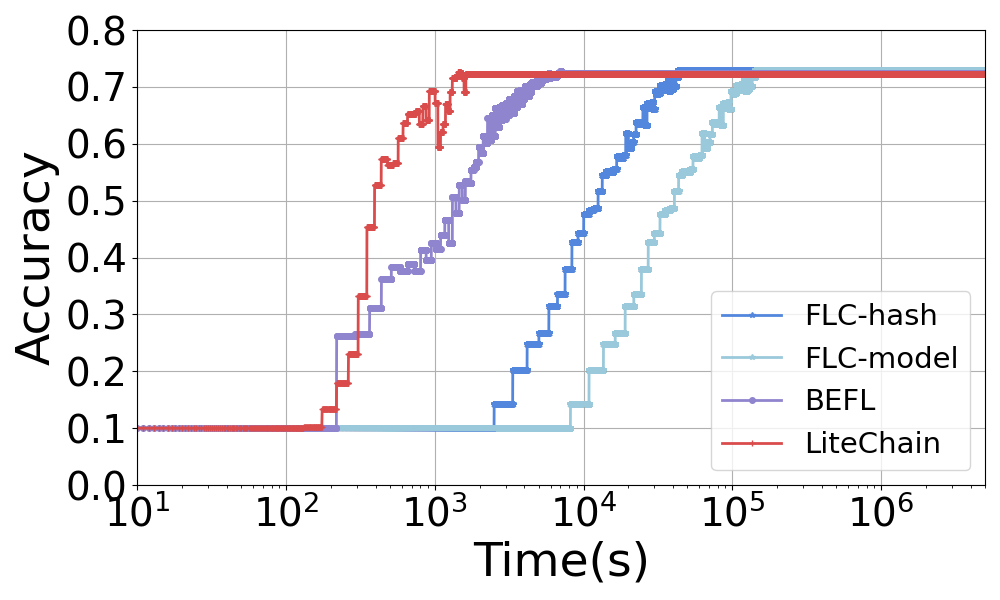}
        \label{s150_02}
    }
    \subfigure[200 devices]
    {
        \includegraphics[scale=0.19]{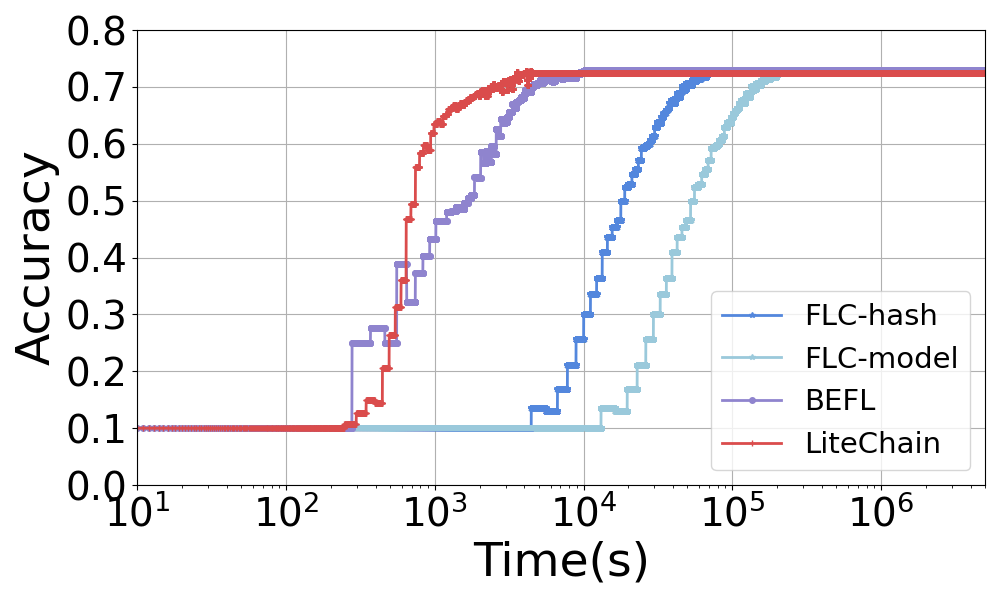}
        \label{s200_02}
    }
    \subfigure[250 devices]
    {
        \includegraphics[scale=0.19]{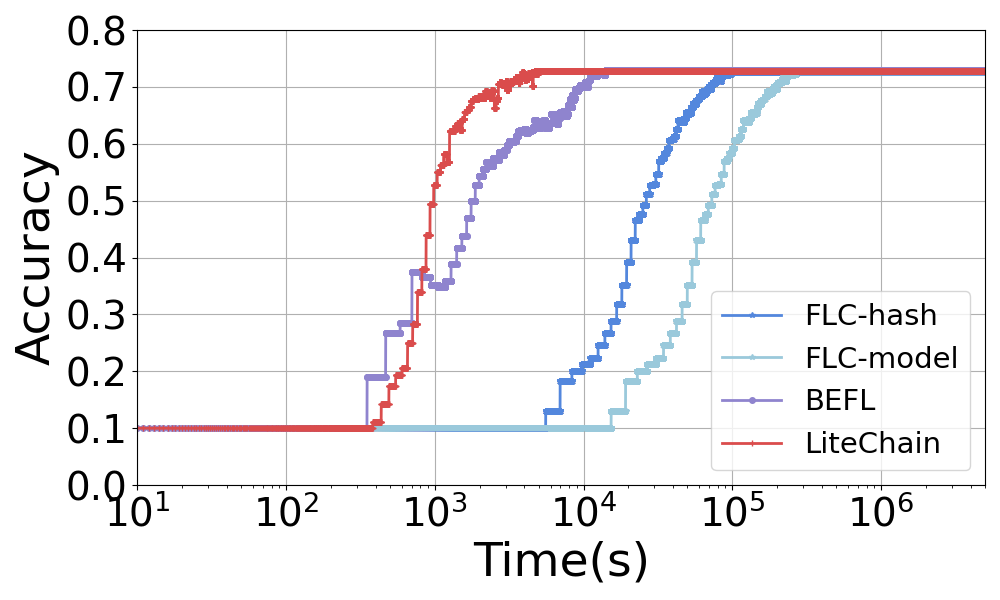}
        \label{s250_02}
    }
    \subfigure[300 devices]
    {
        \includegraphics[scale=0.19]{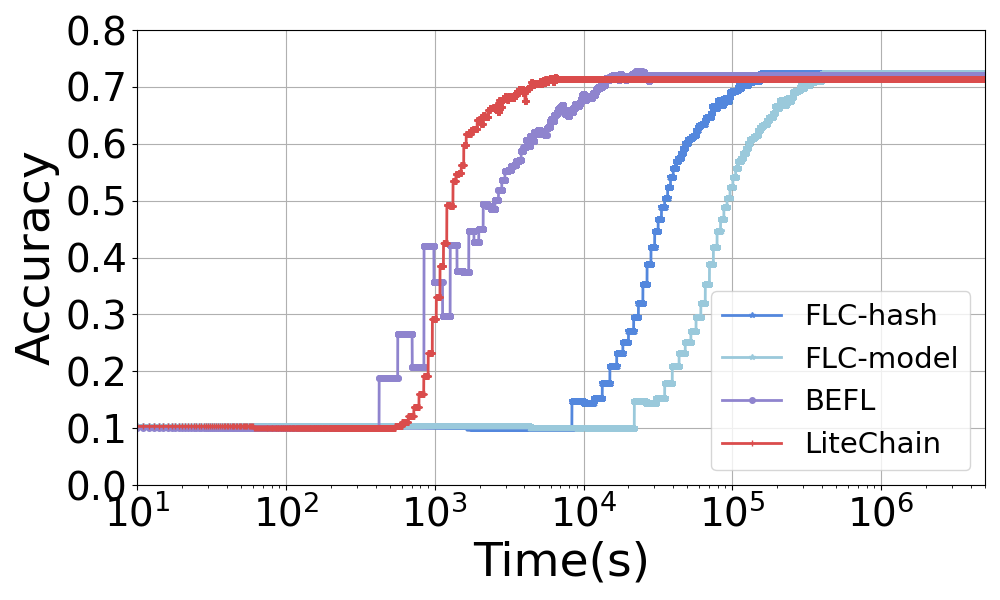}
        \label{s300_02}
    }
    \caption{Accuracy over time (seconds) with non-IID dataset.}
    \label{fig:AccTime02}
\end{figure*}

\textbf{Accuracy over latency.}
Figs. \ref{fig:AccTime5} and \ref{fig:AccTime02} depict the test accuracy over time on IID and non-IID datasets, respectively. We set the minimum achievable accuracy as 0.73 for all the benchmarks, which serves as the termination condition to evaluate the relationship between accuracy and time. The test accuracy of the latest model is evaluated every 1 second. We utilize a log-scale x-axis to display the benchmarks in one figure. 

Figs. \ref{s50_5}$\sim$\ref{s300_5} illustrate the test accuracy over time across 50 to 300 devices on the IID dataset. Initially, the accuracy of BEFL increases quickly, especially with 200-300 devices. As time passes, LiteChain shows a more steady improvement in accuracy and swiftly reaches the termination criterion. LiteChain takes 264 seconds and 3852 seconds to reach the expected accuracy across 50 and 300 devices, respectively, about only 27\% and 33\% as long as the latency of BEFL. LiteChain reaches the termination accuracy with the lowest latency under varying network scales. The results demonstrate the scalability of LiteChain in IID datasets.

To attenuate the benefits of homogeneous local data distribution on convergence rate, Figs. \ref{s50_02}$\sim$\ref{s300_02} present the test accuracy over time on the non-IID datasets across 50 to 300 devices. The training latency of LiteChain in non-IID is about 1.73$\sim$1.83 times of that in the IID dataset. The latency taken by LiteChain with 50 devices and 300 devices is 0.18 and 0.23 time of BEFL’s training latency, respectively. Overall, LiteChain shows robustness to the non-IID training data and maintains a rapid convergence rate as the network scale increases. Incorporating the insights from Fig. \ref{fig:AccTime5}, LiteChain outperforms FLC (FLC-hash and FLC-model) and BEFL in that it reaches the termination condition more rapidly under both IID and non-IID data settings. The results demonstrate LiteChain’s ubiquitous scalability without assuming the training data distributions.

\begin{figure}[ht]
    \centering
    \includegraphics[width = 0.95\linewidth]{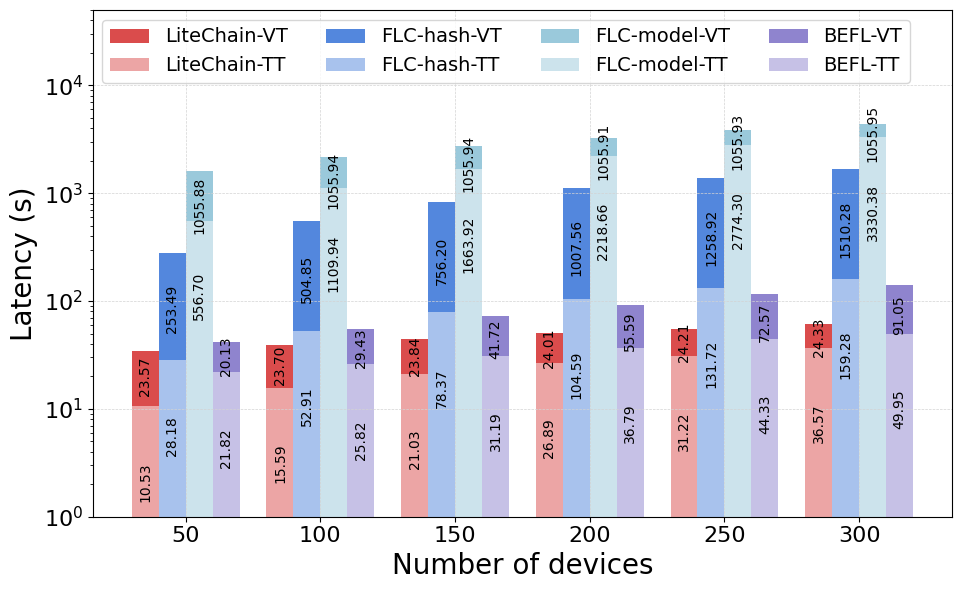}
    \caption{Latency overhead of two specific tasks: Training Task (TT) and Verification Task (VT) in one-round FL training across 50-300 devices.}
    \label{exp:oneround}
\end{figure}

\textbf{Latency overhead.}
To further inspect the latency overhead, we specify the latency overhead of one-round FL training in Fig. \ref{exp:oneround}. Here, the latency of LiteChain is evaluated based on the maximum latency experienced across the network.
The overall latency in one-round training includes two phases: Training Task (TT) and Verification Task (VT).  Herein, TT consists of local training, query records, and aggregation latency, while VT is the latency taken by verifying the updated models.  
As the network scales up, the latency of both TT and VT increases. For TT, LiteChain constructed on the reorganized network requires fewer models in one aggregation step compared to the other two benchmarks. The reduction decreases latency in querying and transmitting the updated models during one-round training.
For VT, the elected committee with fewer committee members in LiteChain necessitates lower communication overhead in consensus models for verification.
LiteChain’s latency is 13.38 seconds and 66.72 seconds less than the latency of BEFL in TT and VT, respectively. As the network scales up, LiteChain maintains a low latency in one-round training.

\subsubsection{Storage Evaluation}

Table \ref{tab:storage} provides the comparison of on-chain storage cost for 1-round and 100-round training across increasing network scales. 
The storage cost of the compared benchmarks is evaluated with the same data storage format to preserve the original data structure for convenience and integrity. To store larger models in blocks with size limitations, the models in FLC-model are split into multiple fragments and stored across multiple blocks. 
For example, a file originally sized at 26.3MB now occupies 147.24MB of storage due to fragmentation and the signatures for multiple blocks. 
The sharply escalating storage burden demonstrates that while the approach may be feasible for short-term applications or tiny model storage, it becomes increasingly unsustainable with training. In this study, LiteChain executes an update consensus mechanism every 20 rounds to periodically clean the stale data. LiteChain not only minimizes storage space per round but also maintains a consistently low storage demand compared to BEFL and FLC-hash.

\begin{table*}[ht]
    \centering
    \renewcommand{\arraystretch}{1.5}
    \caption{The storage cost by Blockchain schemes in different number of devices.}
    \label{tab:storage}
    \scalebox{0.75}{ \begin{tabular}{p{1.5cm}|>{\centering\arraybackslash}p{1.25cm}|>{\centering\arraybackslash}p{1.25cm}|>{\centering\arraybackslash}p{1.25cm}|>{\centering\arraybackslash}p{1.25cm}|>{\centering\arraybackslash}p{1.25cm}|>{\centering\arraybackslash}p{1.25cm}|>{\centering\arraybackslash}p{1.25cm}|>{\centering\arraybackslash}p{1.25cm}|>{\centering\arraybackslash}p{1.25cm}|>{\centering\arraybackslash}p{1.25cm}|>{\centering\arraybackslash}p{1.25cm}|>{\centering\arraybackslash}p{1.25cm}}
    \hline
        \multirow{2}{*}{\textbf{Schemes}} & \multicolumn{6}{c|}{\textbf{1 Round}} & \multicolumn{6}{c}{\textbf{100 Rounds}} \\ 
    \cline{2-13}
         & \textit{50} & \textit{100} & \textit{150} & \textit{200} & \textit{250} & \textit{300} & \textit{50} & \textit{100} & \textit{150} & \textit{200} & \textit{250} & \textit{300} \\
    \hline
         FLC-model & $147.242$MB & $147.335$MB & $147.427$MB & $147.519$MB & $147.612$MB & $147.704$MB & $14.379$GB & $14.388$GB & $14.397$GB & $14.406$GB & $14.415$GB & $14.424$GB\\
    \hline
         FLC-hash & $96.743$KB & $191.303$KB & $285.856$KB &	$380.421$KB & $474.979$KB & $569.536$KB & $9.448$MB & $18.682$MB & $27.916$MB & $37.150$MB & $46.385$MB & $55.619$MB\\
    \hline
         BEFL & $697.097$KB & $697.096$KB & $697.097$KB & $697.096$KB & $697.097$KB & $697.098$KB & $3.480$MB & $3.479$MB & $3.480$MB & $3.480$MB & $3.480$MB & $3.480$MB \\
    \hline
         \textbf{LiteChain} & $29.633$KB & $57.090$KB & $74.407$KB & $86.215$KB & $128.595$KB & $142.400$KB & $0.579$MB & $1.115$MB & $1.453$MB & $1.684$MB & $2.512$MB & $2.781$MB\\
    \hline
    \end{tabular}}
\end{table*}

\subsection{Security Evaluation}
This section evaluates the {consensus security} from simulation and experiments with attackers. We first introduce the security score evaluation based on simulated devices with different reliability ranges. Then, we test the accuracy performance under adversarial settings.

% \subsubsection{\colorbox{yellow}{Consensus Security Simulation}}
\subsubsection{{Consensus Security Simulation}}
\begin{figure}[ht]
    \centering
    \includegraphics[width = 0.8\linewidth]{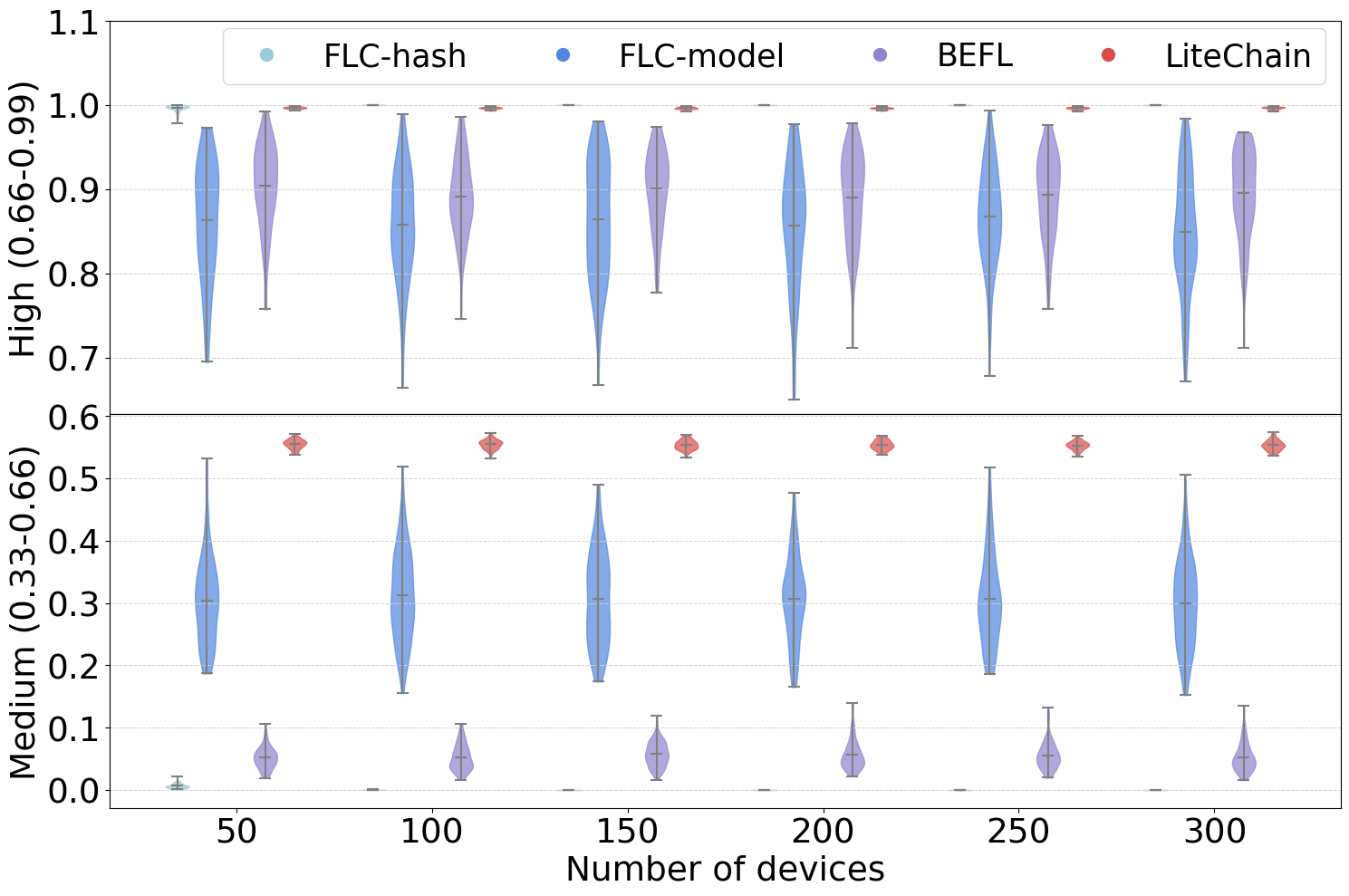}
    \caption{Consensus security scores across 100 times. The generated reliability ranges from top to bottom: High (0.66–0.99), Medium (0.33–0.66).}
    \label{fig:reputation}
\end{figure}
Fig. \ref{fig:reputation} evaluates the consensus security score calculated according to equation (\ref{equ:security}) over 100 experimental networks and visualizes it as violin plots to reflect the score distribution. The reliability (0-1) is generated with two categories: medium (0.33–0.66) and high (0.66–0.99) to test system robustness under varying levels of inherent consensus security risks. Low-reliability scenarios with (0-0.33) are insecure for task execution in the real world, so they are not considered in this study. In the high reliability range (0.66-0.99), FLC-hash achieves the highest consensus security due to a large number of committee members. However, in the medium reliability range (0.33-0.66), the consensus security of FLC-hash is significantly lower when all the devices participate in the consensus mechanism. FLC-model is constrained by broadcast timeout, preventing it from broadcasting the block to all the committee members, leading to a considerable bias in 100 different simulated networks.
LiteChain maintains high consensus security through a selection targeting at those trustworthy devices to serve as the committee members. Despite experiencing a drop in consensus security scores within the low safety probability range, LiteChain's performance remains notably superior to all the other benchmarks.

\subsubsection{Model Protection Performance}
To evaluate the robustness of LiteChain, we consider two attacks to disrupt the training performance. 

\begin{figure}[ht]
    \centering
    \subfigure[50 devices]
    {
        \includegraphics[scale=0.15]{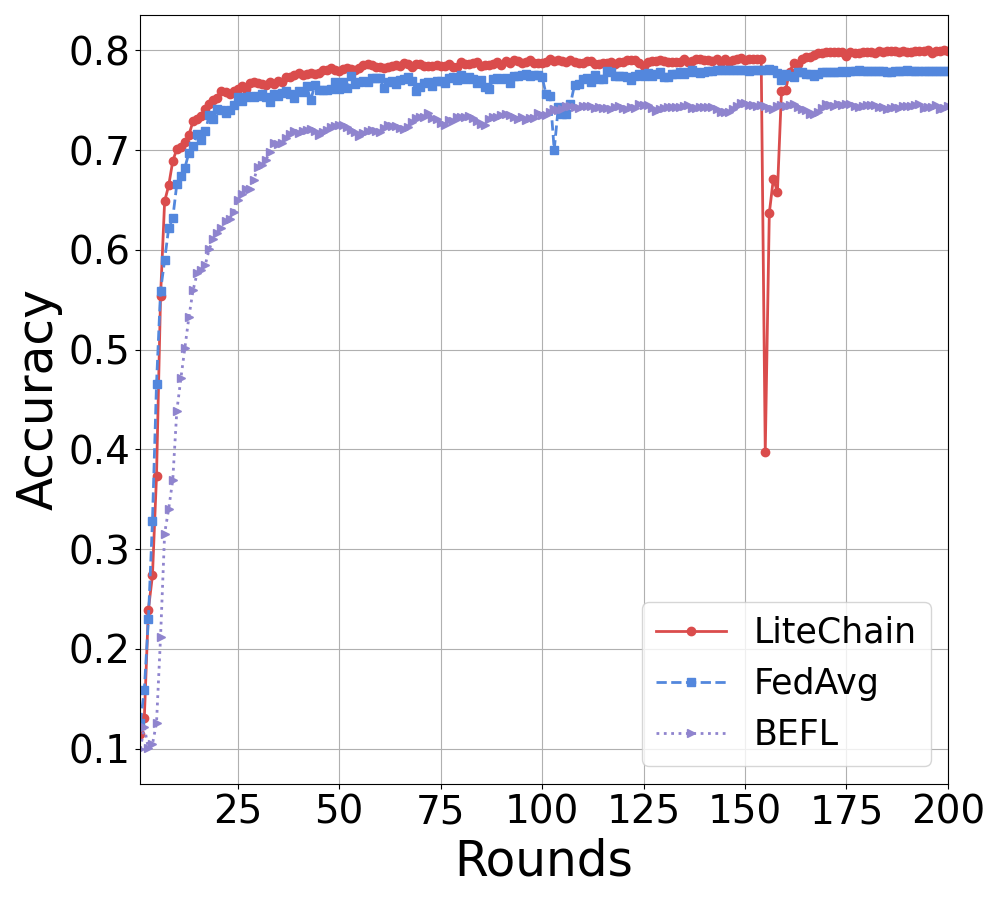}
        \label{a2_50_5}
    }
    \subfigure[300 devices]
    {
        \includegraphics[scale=0.15]{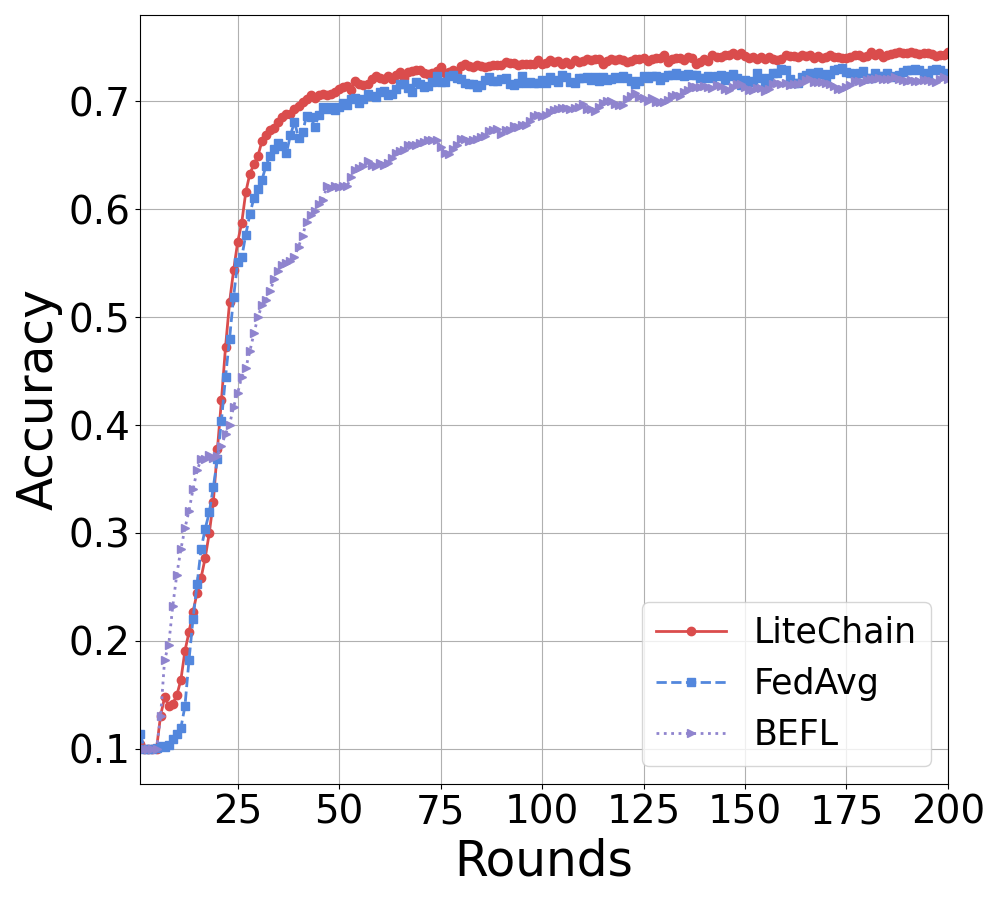}
        \label{a2_s300_5}
    }
    \caption{Accuracy under replay attack on IID dataset (EFL represents BEFL without blockchain empowerment. FedAvg represents FLC without blockchain empowerment).}
    \label{exp:a2_iid}
\end{figure}

\begin{figure}[!ht]
    \centering
    \subfigure[50 devices]
    {
        \includegraphics[scale=0.15]{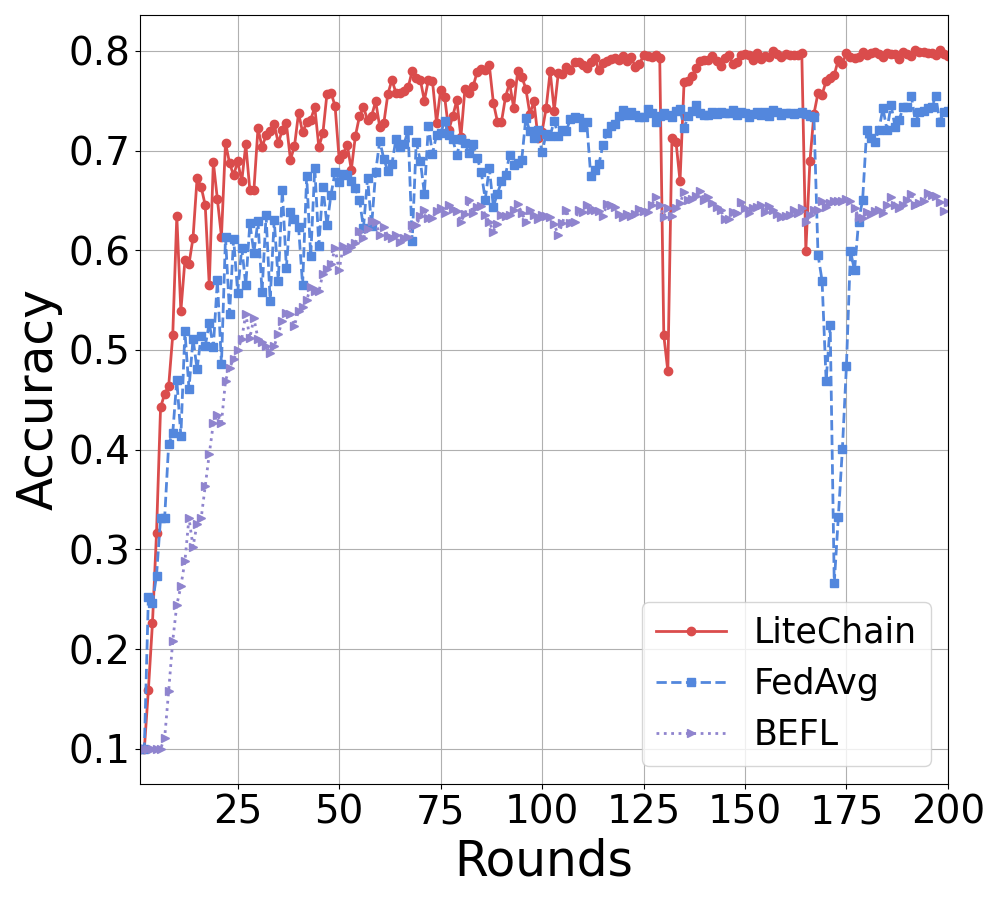}
        \label{a2_s50_02}
    }
    \subfigure[300 devices]
    {
        \includegraphics[scale=0.15]{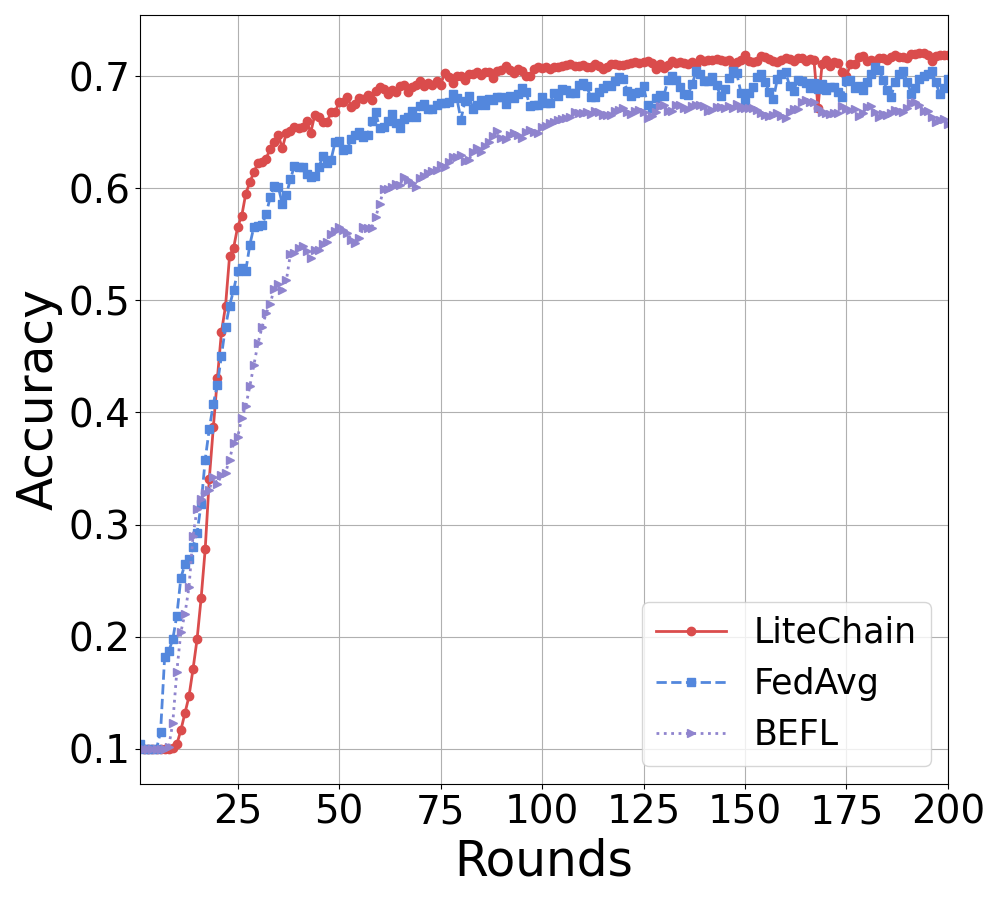}
        \label{a2_s300_02}
    }
    \caption{Accuracy under replay attack with non-IID dataset (EFL represents BEFL without blockchain empowerment. FedAvg represents FLC without blockchain empowerment).}
\label{exp:a2_niid}
\end{figure}

\textbf{Replay attack.} In replay attacks, malicious devices try to obtain more rewards by submitting stale models. Figs. \ref{exp:a2_iid} and \ref{exp:a2_niid} evaluate the training performance under replay attackers with 0.5 attacking rate in IID and non-IID training datasets. With the designed consensus mechanism to track historical transactions, duplicated models can be detected easily by checking model identifiers, timestamps, and other information. Since the benchmarks compared in our study do not mention this function for duplicate detection, we use 'EFL' to denote BEFL without blockchain empowerment and 'FedAvg' to represent FLC without blockchain empowerment for differentiation. According to space constraints, we show the performance results in the smallest and the largest networks with 50 and 300 devices, respectively. More experiment results in the networks across 100 to 250 devices can be found in Appendix G. The experiment results demonstrate that LiteChain exhibits superior performance compared to FedAvg and EFL with both 50 and 300 devices. The experiment results emphasize the notable influence of replay attacks on the convergence of FL training. The converged accuracy of LiteChain is higher than the other benchmarks. Its robustness was particularly evident in Fig. \ref{exp:a2_niid}, where data heterogeneity typically impacts convergence performance. 

\begin{figure}
    \centering
    \includegraphics[width = 0.8\linewidth]{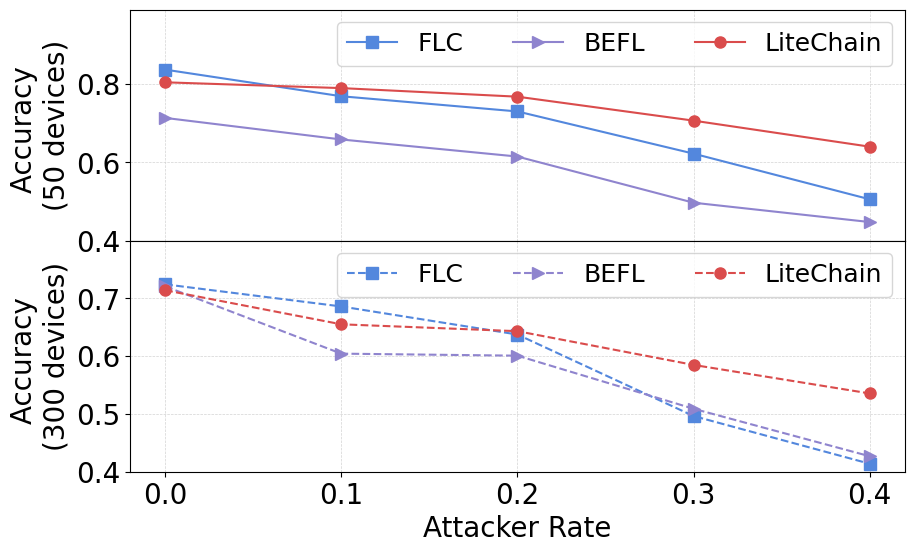}
    \caption{Accuracy under data poisoning attack on IID dataset.}
    \label{fig:a1_1}
\end{figure}

\begin{figure}
    \centering
    \includegraphics[width = 0.8\linewidth]{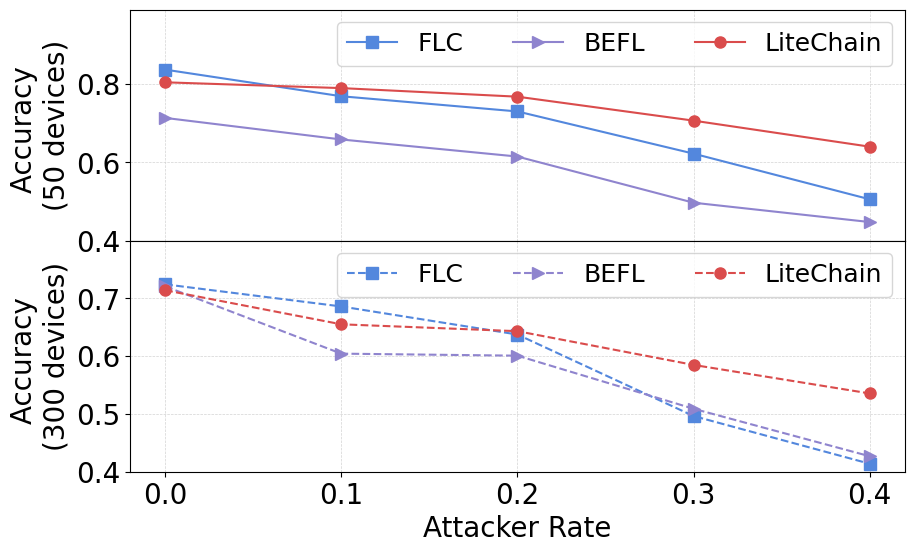}
    \caption{Accuracy under data poisoning attack on non-IID dataset.}
    \label{fig:a1_2}
\end{figure}

\textbf{Data poisoning attack.} By executing label flipping with $(\text{label}+1)\%10$, the attackers inject mislabeling data into the training phase to degrade the test accuracy. Figs. \ref{fig:a1_1} and \ref{fig:a1_2} show the evaluation of FL models' resilience to poison attacks across 50 and 300 devices in IID and non-IID datasets, respectively. The attacker rate denotes the proportion of participants acting maliciously.
As the attacker rate increases, the accuracy of all the models decreases as expected since more corrupted data is introduced into the system. BEFL performs well under lower attacker rates but loses accuracy sharply as the attacker rate increases. {LiteChain implementing an accuracy threshold checks the quality of the uploaded model before the first aggregation to prevent adversely impacting the quality of the aggregated model. We consider a network with 300 devices to evaluate the model accuracy, varying the number of attackers from 0 to 40\%. In the IID dataset, as shown in Fig. {\ref{fig:a1_1}}, LiteChain's accuracy only decreases by 0.07, whereas the accuracy of BEFL and FLC drop by 0.40 and 0.23, respectively. In non-IID datasets {\ref{fig:a1_2}}, LiteChain's accuracy decreases by 0.17, while the accuracy of BEFL and FLC decrease by 0.29 and 0.31, respectively. } 
LiteChain is more effective at mitigating the impact of malicious inputs than FLC and BEFL. 
Experiment results demonstrate that LiteChain with hierarchical training maintains a higher fault tolerance and stability across networks with various scales. Furthermore, LiteChain is compatible with other secure FL aggregation methods under different objectives.

% \subsection{Discussion}

% \begin{itemize}
%     \item Dynamic state: network, devices
%     \item Traffic congestion
%     \item Other attacks
% \end{itemize}

\section{Conclusion}
In this study, we introduced LiteChain, a lightweight blockchain framework for scalable and verifiable FL in MENs. To establish the communication-efficient LiteChain with optimized computation utilization, we designed a distributed clustering algorithm by trading off between latency and consensus security to reorganize MENs and elect committee members to construct committees. We presented a novel consensus security metric and utilized a discrete Fourier transform for rapid calculation. The reliability of the training process was ensured by integrating intra-cluster training with off-chain {verification} and inter-cluster training with on-chain CBFT consensus. Furthermore, to minimize storage overhead while preserving power centralization, we introduced an update consensus mechanism to clean stale storage and update committee members periodically. The theoretical analysis and the experiment results demonstrated the cost-efficiency and robustness of LiteChain, highlighting its potential for scalable implementation in MENs.

{Nevertheless, LiteChain requires the entire model for verification, which increases the risk of privacy leakage through inference attacks by curious committee members. Differential privacy and homomorphic encryption are two common strategies to address this issue. However, traditional differential privacy would reduce training effectiveness and model verifiability. Homomorphic encryption is difficult to apply on a large scale due to significant performance overhead and computational complexity. In the future, we plan to enhance our work by designing effective lightweight verification mechanisms robust to inference attacks.}

\ifCLASSOPTIONcompsoc
  \section*{Acknowledgments}
\else
  \section*{Acknowledgment}
\fi
This research was supported by the UGC General Research Funds No. 17203320 and No. 17209822 from Hong Kong.

\ifCLASSOPTIONcaptionsoff
  \newpage
\fi

\bibliographystyle{ieeetr}
\bibliography{Refe}

\begin{IEEEbiography}[{\includegraphics[width=1in,height=1.25in,clip,keepaspectratio]{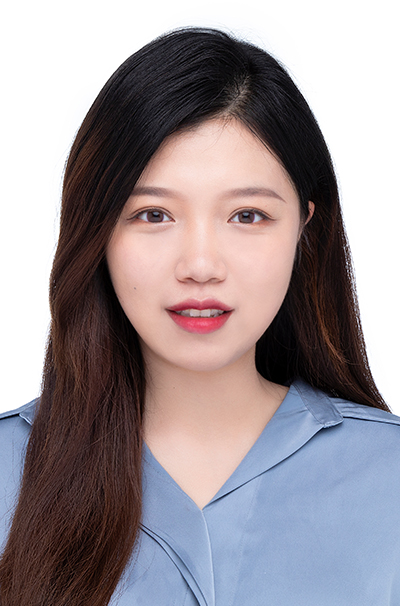}}]{Handi Chen} 
    received the M.Sc. degrees in Network Engineering in 2022 from the Dalian University of Technology, Dalian, China. She is currently working toward the Ph.D. degree in Department of Electrical and Electronic Engineering, the University of Hong Kong, Hong Kong, China. Her research interests include wireless communication, mobile edge computing and privacy preservation.
\end{IEEEbiography}

\begin{IEEEbiography}[{\includegraphics[width=1in,height=1.25in,clip,keepaspectratio]{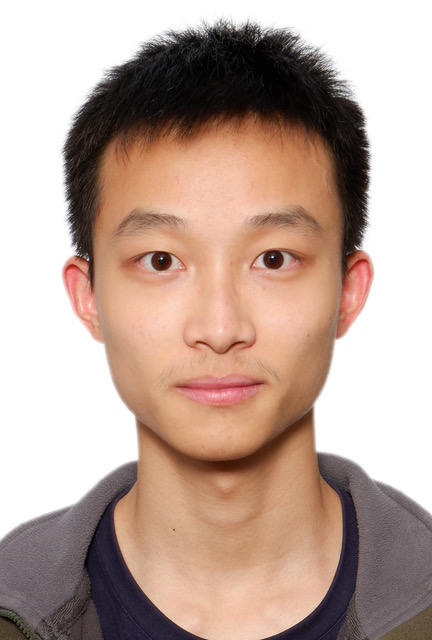}}]{Zhou Rui} 
    received his BSc degree in 2022 from the University of Hong Kong, Hong Kong, China. He is now a PhD student at HKU Internet-of-Things Lab. His research interest centres around distributed optimization, privacy-preserving machine learning, and resource-constrained machine learning.
\end{IEEEbiography}

\begin{IEEEbiography}[{\includegraphics[width=1in,height=1.25in,clip,keepaspectratio]{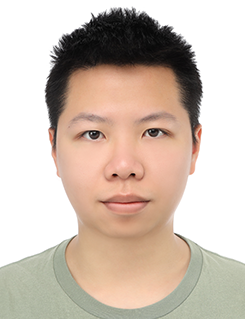}}]{Yun-Hin Chan} 
    received a B.Eng. of Software Engineering from Sun Yat-sen University in 2019. His work is focused on how to solve practical challenges in federated learning, such as communication efficiency and asynchronous algorithms. His research interests include deep learning, asynchronous and distributed optimization, federated learning, knowledge distillation and transfer learning.
\end{IEEEbiography}

\begin{IEEEbiography}[{\includegraphics[width=1in,height=1.25in,clip,keepaspectratio]{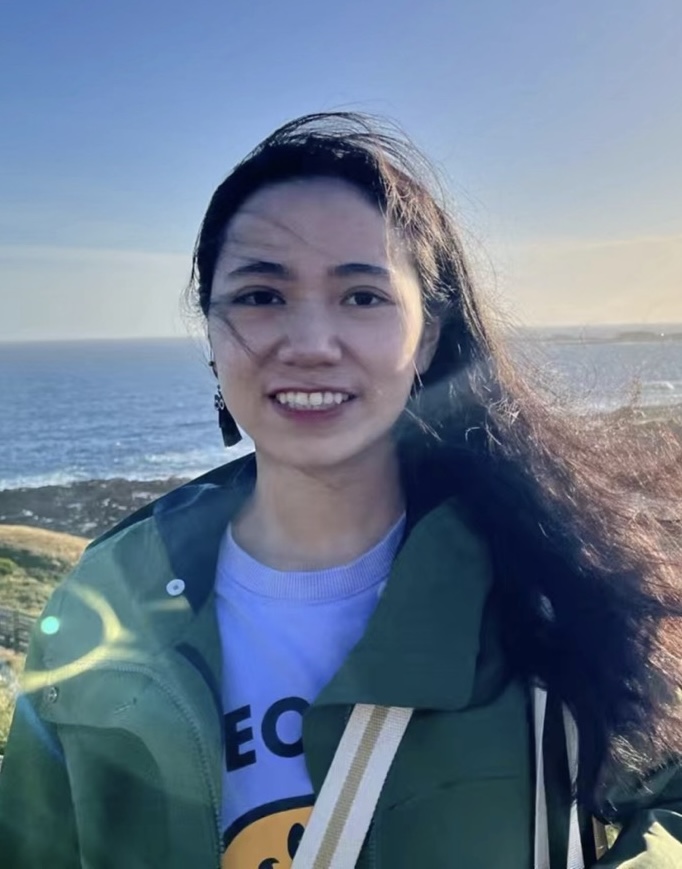}}]{Zhihan Jiang} 
    received the B.E. and M.E. degrees in computer science and technology from Xiamen University, Xiamen, China, in 2018 and 2021, respectively. She is currently pursuing the Ph.D. degree with the Department of Electrical and Electronic Engineering, The University of Hong Kong. Her research interests include Data Analytics and Visualization, Ubiquitous Computing, and Mobile Computing.
\end{IEEEbiography}

\begin{IEEEbiography}[{\includegraphics[width=1in,height=1.25in,clip,keepaspectratio]{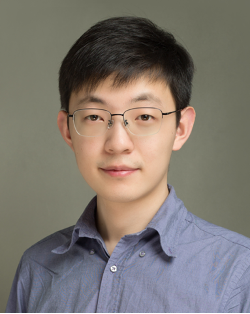}}]{Xianhan Chen} 
    received the B.Eng. degree in electronic information from Southwest Jiaotong University in 2017, and the Ph.D. degree in electrical and computer engineering from the University of Florida in 2022. He is currently an assistant professor at the Department of Electrical and Electronic Engineering, the University of Hong Kong. He serves as an Associate Editor of ACM Computing Surveys. He received the 2022 ECE graduate excellence award for research from the University of Florida. His research interests include wireless networking, edge intelligence, and machine learning.
\end{IEEEbiography}

\begin{IEEEbiography}[{\includegraphics[width=1in,height=1.25in,clip,keepaspectratio]{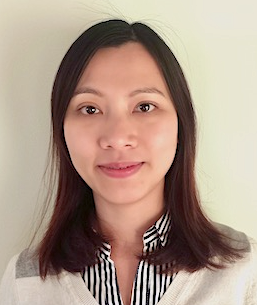}}]{Edith C.H. Ngai} 
    is currently an Associate Professor in the Department of Electrical and Electronic Engineering, The University of Hong Kong. Before joining HKU in 2020, she was an Associate Professor in the Department of Information Technology, Uppsala University, Sweden. Her research interests include Internet-of-Things, edge intelligence, smart cities, and smart health. She was a VINNMER Fellow (2009) awarded by the Swedish Governmental Research Funding Agency VINNOVA. Her co-authored papers received a Best Paper Award in QShine 2023 and Best Paper Runner-Up Awards in ACM/IEEE IPSN 2013 and IEEE IWQoS 2010. She was an Area Editor of IEEE Internet of Things Journal from 2020 to 2022. She is currently an Associate Editor in IEEE Transactions of Mobile Computing, IEEE Transactions of Industrial Informatics, Ad Hoc Networks, and Computer Networks. She has served as a program chair in IEEE ISSNIP 2015, IEEE GreenCom 2022, and IEEE/ACM IWQoS 2024. She received a Meta Policy Research Award in Asia Pacific in 2022. She was selected as one of the N²Women Stars in Computer Networking and Communications in 2022. She is a Distinguished Lecturer in IEEE Communication Society in 2023-2024.
\end{IEEEbiography}

\newpage

\title{{LiteChain}: A Lite Blockchain for Verifiable and Scalable Federated Learning in Massive Edge Network}

\author{Handi~Chen,
        Rui~Zhou,
        Yun-Hin~Chan,
        Zhihan~Jiang,~\IEEEmembership{Graduate Student Member,~IEEE,} 
        Xianhao~Chen,~\IEEEmembership{Member,~IEEE,} 
        and~Edith~C.~H.~Ngai,~\IEEEmembership{Senior Member,~IEEE}}

\markboth{IEEE TRANSACTIONS ON MOBILE COMPUTING}%
{Chen \MakeLowercase{\textit{et al.}}: {LiteChain}: A Lite Blockchain for Verifiable and Scalable Federated Learning in Massive Edge Network}
\maketitle

\begin{appendices}
\section{Proof of \textbf{Theorem 1}}
\begin{proof}
    We consider the system with a validator set denoted as $\mathcal{V}$. In Byzantine Fault Tolerance (BFT)-based consensus mechanism, we consider $m$ malicious devices. Validator $j$'s reliability is denoted as $p_j$, while the unreliability is $(1-p_j)$. There are $\binom{K}{m}$ combinations of $m$ malicious devices in $K$ validators. The probability of success consensus for each malicious device combination can be calculated as $\prod_{j\in\mathcal{V}_{m'}}(1-p_j)\prod_{j\in\mathcal{V}_{m'}^c}p_j$, where $\mathcal{V}_{m'}$ and $\mathcal{V}_{m'}^c$ are sets of malicious nodes and normal nodes (the complement of malicious nodes), respectively. The probability of successful consensus, defined as system security, can be viewed as the sum of the probabilities of all these combinations. That is,
    \begin{equation}
        S_m = \sum_{{m'}=1}^{\binom{K}{m}}\prod_{j\in\mathcal{V}_{m'}}p_j\prod_{j\in\mathcal{V}_{m'}^c}(1-p_j).
    \end{equation}
    Under BFT, the number of malicious nodes needs to satisfy $\vert\mathcal{V}_{m'}\vert\le\lfloor\frac{K-1}{3}\rfloor$. Therefore, the system security can be represented as follows:
    \begin{align}
        \begin{split}
            S &= \sum_{m=0}^{\lfloor\frac{K-1}{3}\rfloor}S_m\\
            & = \sum_{m=0}^{\lfloor\frac{K-1}{3}\rfloor}\sum_{{m'}=1}^{\binom{K}{m}}\prod_{j\in\mathcal{V}_{m'}}p_j\prod_{j\in\mathcal{V}_{m'}^c}(1-p_j).
        \end{split}
    \end{align}
    The proof is completed.
\end{proof}

\section{Proof of \textbf{Theorem 2.}}
\begin{proof}
    In LiteChain, each device training round consists of two tasks: FL training and Blockchain verification. The constant size of the model during training is denoted as $\Lambda^{size}$. The expected size of the broadcast message during blockchain verification is denoted as $\bar{B}^{size}$.

    \textbf{One-tier network.} Devices in the P2P FL training broadcast the entire model to other devices. The communication complexity expectation of device $i$ is $N(N-1)\Lambda^{size}$.
    For $N$ validators, the communication complexity expectation based on the proposed Comprehensive BFT (CBFT) can be summarized as:
    $[2(N-1)+(N-1)^2+N^2]\bar{B}^{size}$. After consensus, the communication complexity expectation of block synchronous is $(N-1)\bar{B}^{size}$.
    To summarize, the communication complexity expectation of one-tier blockchain-empowered FL can be calculated as follows:
    \begin{equation}
        \mathbb{E}(\mathcal{C}^o) = (N^2-N)\Lambda^{size}+(2N^2+N-2)\bar{B}^{size}.
    \end{equation}

    \textbf{Optimized network.} The network is optimized into $K$ clusters, the number of devices in cluster $k$ is denoted as $n_k$, and $\mathbb{E}(n_k)=N/K$. The number of validators is $K$. The communication complexity expectation of each device in FL training includes two processes, i.e., sending the local model to the validator and receiving the model from the validator. The communication complexity expectation of FL training is denoted as $2N\Lambda^{size}/K$. For $K$ validators, the communication complexity expectation of CBFT and block synchronization is denoted as $\left[2(K-1)+(K-1)^2+K^2\right]\bar{B}^{size}$ and $(K-1)\bar{B}^{size}$, respectively. Therefore, the communication complexity expectation of LiteChain with an optimized network can be represented as follows:
    \begin{equation}
        \mathbb{E}(\mathcal{C}^{lc}) = \frac{2N\Lambda^{size}}{K}+(2K^2+K-2)\bar{B}^{size}.
    \end{equation}

    To summarize, the reduced communication complexity after network optimization can be represented as follows:
    \begin{equation}
        \begin{split}
            \mathbb{E}(\Delta\mathcal{C})=&\Lambda^{size}(N^2-N-\frac{2N}{K})+\bar{B}^{size}(2N^2+N-\\
            &2K^2-K).
        \end{split}
    \end{equation}
    Under BFT, we have $K\ge4$. Thus, the maximum reduced communication complexity after network optimization is 
    \begin{equation}
        \begin{split}
            \max\mathbb{E}(\Delta\mathcal{C})=&\Lambda^{size}(N^2-\frac{3N}{2})+\bar{B}^{size}(2N^2+N-36).
        \end{split}
    \end{equation}
    When $N\ge4$, the maximum expected reduction will be greater than 0. As $N$ increases, the advantages of our structure in reducing communication complexity become increasingly pronounced. The proof is completed.
\end{proof}

\section{Proof of \textbf{Theorem 3}}
\begin{proof}
The social welfare of a partition is defined as $R(\mathbf{K}):=\sum_{k\in K}u(\mathcal{K}_k)$. After executing $t$-th decision-making iterations, we obtain the corresponding partition $\mathbf{K}^{(t)}$ and social welfare $R(\mathbf{K}^{(t)})$. {In each iteration, the node with the fewest visits is selected first to determine the switch operation within a cluster. Hence, the order of the switch operations remains unchanged under the same network conditions.}
Since only switch operations with strictly positive gain (i.e., $G(\sigma(\cdot))>0$) are considered, $R(\mathbf{K}^{(t)})>R(\mathbf{K}^{(t-1)})$ always holds. After any switch operation, the social welfare system strictly increases, and the same partition can never be visited twice. {Furthermore, since there is a finite number of partitions, the algorithm converges to a unique partition result after a finite number of iterations in a fixed order.} The proof is completed.
\end{proof}

\section{Proof of \textbf{Lemma 1}}
\begin{proof}
% \ref{ass1} --> 1
    According to Assumption 1, ${F}$ satisfies $\mathcal{L}$-smooth, we have:
    \begin{align}
        \begin{split}
            &\mathbb{E}\left[{F}(w_{\tau,\phi})\right]-\mathbb{E}\left[{F}(w_{\tau,\phi-1})\right]\\
        \le&\mathbb{E}\left[\langle\nabla{F}(w_{\tau,\phi-1}),w_{\tau,\phi-1}-w_{\tau,\phi}\rangle\right]+\\
        &\frac{\mathcal{L}}{2}\mathbb{E}\left[\Vert{w_{\tau,\phi-1}}-w_{\tau,\phi}\Vert^2\right]\\
        \le&-\eta\mathbb{E}\left[\langle\nabla{F}(w_{\tau,\phi-1}),\nabla{f}(w_{\tau,\phi-1})\rangle\right]+\\
        &\frac{\mathcal{L}\eta^2\mathbb{E}\left[\Vert{\nabla{f}(w_{\tau,\phi-1})}\Vert^2\right]}{2}.
        \end{split}
    \end{align}
    Herein, the term $\mathbb{E}\left[{\nabla}F(w_{\tau,\phi-1})\right]$ can be rewritten as $\mathbb{E}\left[\sum_{k=1}^{K}\sum_{i=1}^{n_k}\nabla{f(w_{i,k;\tau,\phi-1})}\right]$, which simplifies to $\sum_{k=1}^{K}\sum_{i=1}^{n_k}\mathbb{E}[f(w_{i,k;\tau,\phi-1})]$. To simplify, the subscripts of clusters and devices ($k$ and $i$) are dropped. Thus, we can rearrange the equation to derive the upper bound of gradients from $\phi-1$ to $\phi$ as follows:
    \begin{align}\label{ie:onceupdate}
        \begin{split} 
        &\mathbb{E}\left[{F}(w_{\tau,\phi})\right]-\mathbb{E}\left[{F}(w_{\tau,\phi-1})\right]\\
        \le&-\eta\mathbb{E}[\Vert\nabla{F}_{w_{\tau,\phi-1}}\Vert^2]+\frac{\mathcal{L}\eta^2Q_1}{2}.
        \end{split}
    \end{align}
    To telescope inequality (\ref{ie:onceupdate}) from step $0$ to $\Phi$, we can obtain:
    \begin{align}
        \begin{split}
            \mathbb{E}\left[F(w_{\tau,\Phi})\right]\le&\mathbb{E}\left[F(w_{\tau,0})\right]-\eta\sum_{\phi=1}^{\Phi}\mathbb{E}[\Vert\nabla{F}(w_{\tau,\phi-1})\Vert^2]\\
            &+\frac{\mathcal{L}\eta^2Q_1\Phi}{2}.
        \end{split}
    \end{align}
    The proof is completed.
\end{proof}

\section{Proof of \textbf{Lemma 2}}
\begin{proof}
% ass3 --> 3
    First, under Assumption 3, there are at most $\mathcal{T}$ steps between $\tau$ and $t-1$. We have:
    \begin{align}
        \begin{split}
            &\mathbb{E}\left[\Vert{w}_{k;\tau} - w_{k;t-1}\Vert^2\right]\\
            % \le&\mathbb{E}[\Vert(w_\tau-w_{\tau-1})+\dots+(w_{t}-w_{t-1})\Vert^2]\\
            =&\mathbb{E}\left[\Vert\sum_{\Delta{\tau}=0}^{t-\tau-2}(w_{k;\tau+\Delta{\tau}}-w_{k;\tau+\Delta{\tau}+1})\Vert^2\right].
        \end{split}
    \end{align}
    According to Cauchy's inequality and $\mathbb{E}[{w_{k;\tau+\Delta{\tau},\Phi}}]=\mathbb{E}[{w_{k;\tau+\Delta{\tau}+1,0}}]$, we have:
    \begin{align}
    \label{ie:upper}
        \begin{split}
            &\mathbb{E}\left[\Vert{w}_{k;\tau}-w_{k;t-1}\Vert^2\right]\\
            \le&(t-\tau-2)\sum_{\Delta{t}=0}^{t-\tau-2}\mathbb{E}[\Vert{s(w_{k;\tau+\Delta{\tau},0}-w_{k;\tau+\Delta{\tau},\Phi}})\Vert^2].
        \end{split}
    \end{align}
    Under Assumptions 2 and 3 and equation (15), inequality (\ref{ie:upper}) can be bounded as follows:
    \begin{align}
        \mathbb{E}\left[\Vert{w}_{\tau} - w_{t-1}\Vert^2\right]\le{\mathcal{T}^2}s^2\Phi^2\eta^2\sqrt{Q_1}.
    \end{align}
    Applying the same argument in $\mathbb{E}\left[\Vert{w}_{\tau} - w_{t-1}\Vert\right]$, we can obtain:
    \begin{align}
        \begin{split}
            &\mathbb{E}\left[\Vert{w}_{k;\tau} - w_{k;t-1}\Vert\right]\\
            \le&\sum_{\Delta{\tau}=0}^{t-\tau-2}\mathbb{E}\left[s(w_{k;\tau+\Delta{\tau},0}-w_{k;\tau+\Delta{\tau},\Phi})\right]\\
            \le&\mathcal{T}s\Phi\eta\sqrt{Q_1}.
        \end{split}
    \end{align}
    The proof is completed.
\end{proof}

\section{Proof of Theorem 4}
\begin{proof}
    According to $\mathcal{L}$-smooth assumption, we have:
    \begin{align}
        \begin{split}
            &\mathbb{E}\left[F(w_t)-F({w_{t-1}})\right]\\
            \le&\mathbb{E}\left[G(w_t)-F(w_{t-1})\right]\\
            \le&\mathbb{E}\left[G(w_{t-1}-s_{k;t-1}w_{k;t-1}+s_{k;\tau}w_{k;\tau})-F(w_{t-1})\right]\\
            \le&\mathbb{E} \left[F(w_{t-1})+\frac{\mu}{2}\Vert w_{t-1}\Vert^2 - s_{k;t-1}F(w_{k;t-1})-\right.\\
            &\left.\frac{\mu}{2}\Vert w_{t-1}\Vert^2 + s_{k;t}F(w_{k;t})+\frac{\mu}{2}\Vert w_{k;t}\Vert^2 -F(w_{t-1})\right] \\
            =&\mathbb{E}\left[s_{k;t}F(w_{k;t})-s_{k;t-1}F(w_{k;t-1})\right]+\frac{\mu}{2}\mathbb{E}\left[\Vert w_{t-1}\Vert^2\right.\\
            &\left.-\Vert w_{k;t-1}\Vert^2+\Vert w_{k;t}\Vert^2\right].
        \end{split}
    \end{align}
    That means the contribution gap of cluster $k$ between step $t-1$ and $t$. Since the computational capability in cluster $k$ will not change with the training step, the staleness of each training step remains constant, i.e., $s_{k;t-1}=s_{k;t}$. To simplify, we drop the subscript for staleness and obtain:
    \begin{align}
        \begin{split}
            &\mathbb{E}\left[F(w_t)-F(w_{t-1})\right]\\
            \le&\mathbb{E}\left[s\left(F(w_{k;t})-F(w_{k;\tau})+F(w_{k;\tau})-F(w_{k;t-1})\right)\right] \\
            &+ \frac{\mu}{2}\mathbb{E}\left[\Vert w_{t-1}\Vert^2-\Vert w_{k;t-1}\Vert^2+\Vert w_{k;t}\Vert^2\right].
        \end{split}
    \end{align}
    
    This inequality includes three terms, i.e., $\mathbb{E}\left[s(F(w_{k;t})-F(w_{k;\tau}))\right]$, $\mathbb{E}\left[s(F(w_{k;\tau})-F(w_{k;t-1}))\right]$ and $\frac{\mu}{2}\mathbb{E}\left[\Vert w_{t-1}\Vert^2-\Vert w_{k;t-1}\Vert^2+\Vert w_{k;t}\Vert^2\right]$. 

    For $s(F(w_{k;t})-F(w_{k;\tau}))$, the model at step $t$ provided by cluster $k$ is synchronously aggregated after local training. By integrating Lemma 1, we can obtain:
    % \ref{le:lu} --> 1
    \begin{align}
    \label{ie:r1}
    \begin{split}
        &\mathbb{E}[F(w_{k;t}) - F(w_{k;\tau})]\\
        =&\sum_{\phi=0}^{\Phi} \mathbb{E}\left[F\left(\frac{\sum_{i=1}^{n_k}\vert{D_i}\vert w_{i,k;\tau,\phi}}{\sum_{i=1}^{n_k}\vert{D_i}\vert}\right)\right] - \mathbb{E}\left[F(w_{k;\tau,0})\right] \\
    \leq& -\eta \sum_{\phi=0}^{\Phi} \mathbb{E}\left[\|\nabla F(w_{\tau,\phi})\|^2\right] + \frac{\mathcal{L}}{2}\eta^2Q_1{\Phi} \\
    \end{split}
    \end{align}
    
    For $F(w_{k;\tau})-F(w_{k;t-1})$, we have:
    \begin{align}
    \label{ie:r2}
        \begin{split}
            &F(w_{k;\tau})-F(w_{k;t-1})\\
            \le&\Vert\nabla{F}(w_{k;t-1})\Vert\Vert w_{k;\tau}-w_{k;t-1}\Vert+\\
            &\frac{\mathcal{L}}{2}\Vert{w_{k;\tau}-w_{k;t-1}}\Vert^2.\\
            % \le&\Vert\nabla{F(w_{t-1})}\Vert\Vert{w_{\tau}-w_{t-1}}\Vert+\frac{\mathcal{L}}{2}\Vert{w_{\tau}-w_{t-1}}\Vert^2\\
        \end{split}
    \end{align}
    According to the result of Lemma 2, we have:
    % \ref{le:st} --> 2
    \begin{align}
        &F(w_{k;\tau})-F(w_{k;t-1})\\
        \le&\mathcal{T}s\Phi\eta\sqrt{Q_1Q_2}+\frac{\mathcal{L}\mathcal{T}^2s^2\Phi^2\eta^2Q_1}{2}.
    \end{align}
    For $\frac{\mu}{2}\mathbb{E}\left[\Vert w_{t-1}\Vert^2-\Vert w_{k;t-1}\Vert^2+\Vert w_{k;t}\Vert^2\right]$, under Assumption 2, we can obtain :
    % we can obtain the upper bound of model from $w_0$:
    \begin{align}\label{ie:r22}
        \begin{split}
            \frac{\mu}{2}\mathbb{E}\left[\Vert w_{t-1}\Vert^2-\Vert w_{k;t-1}\Vert^2+\Vert w_{k;t}\Vert^2\right]\le \mu W
        \end{split}
    \end{align}
    % Therefore, 
    % \begin{align}
    %     \begin{split}
    %         &\frac{\mu}{2}\mathbb{E}\left[\Vert w_{t-1}\Vert^2-\Vert w_{t-1;k}\Vert^2+\Vert w_{t;k}\Vert^2\right]\\
    %         \le&\mu(\eta^2{Z}^2 +2(T^*)\eta^2{Z}\sqrt{Q_2}+{T^*}^2\eta^2Q_2)
    %     \end{split}
    % \end{align}
    By integrating inequalities (\ref{ie:r1}), (\ref{ie:r2}) and (\ref{ie:r22}), we have:
    \begin{align}
    \label{ie:r3}
        \begin{split}
            &\mathbb{E}\left[F(w_{t+1}) - F(w_{t})\right] \\
            \leq& -s\eta \sum_{\phi=0}^{\Phi} \mathbb{E}\left[\|\nabla F(w_{\tau,\phi})\|^2\right] + \frac{\mathcal{L}s\eta^2Q_1\Phi}{2}\\
            &+ \mathcal{T}s^2\Phi\eta\sqrt{Q_1Q_2}+\frac{\mathcal{L}\mathcal{T}^2s^3\Phi^2\eta^2Q_1}{2}+\mu W.
        \end{split}
    \end{align}
    We can rearrange inequality (\ref{ie:r3}), and obtain:
    \begin{align}
    \label{ie:gradient}
        \begin{split}
            &\sum_{\phi=0}^{\Phi}\mathbb{E}\left[\Vert \nabla F(w_{\tau,\phi})\Vert^2\right]\\
            \le&\frac{\mathbb{E}\left[F(w_{t})-F(w_{t+1})\right]}{s\eta}+ \frac{\mathcal{L}\eta Q_1\Phi}{2}\\
            &+ \mathcal{T}s\Phi\sqrt{Q_1Q_2}+\frac{\mathcal{L}\mathcal{T}^2s^2\Phi^2\eta Q_1}{2}+\frac{\mu W}{s\eta}.
        \end{split}
    \end{align}
    Therefore, we can obtain the upper bound by telescoping (\ref{ie:gradient}) from $t=0$ to $T$ as:
    \begin{align}
        \begin{split}
            &\min_{t=0,\cdots,T }\mathbb{E}\left[\|\nabla F(w_{t})\|^2\right]\\
            \leq &\frac{1}{\sum_{t=1}^{T} \Phi}\sum_{t=1}^{T}\sum_{\phi=1}^{\Phi}\|\nabla F(w_{\tau,\phi})\|^2 \\
            \leq &\frac{\mathbb{E}\left[F(w_{0}) - F(w_{T})\right]}{s \eta T \Phi} +\frac{\mathcal{L}\eta Q_1}{2}+\mathcal{T}s\sqrt{Q_1Q_2}\\
            &+\frac{\mathcal{L}\mathcal{T}^2s^2\Phi\eta Q_1}{2}+\frac{\mu W}{s\eta\Phi}.
            % &\eta\left(\frac{\mathcal{L}sQ_1}{2}+\frac{\mathcal{L}sQ_1\mathcal{T}^2\Phi}{2}+\frac{\mu Z^2}{s\Phi}+\right.\\
            % &\left.\frac{2\mu T^*Z\sqrt{Q_2}}{s\Phi}+\frac{\mu{T^*}^2Q_2}{s\Phi}\right)
            % &\eta\left[\frac{\mathcal{L}sQ_1(1+\mathcal{T}^2\Phi)}{2}+\frac{\mu(\Vert{{Z}}\Vert^2 +2T^*{Z}\sqrt{Q_2}+{T^*}^2 Q_2)}{s\Phi}\right]\\
            % &\frac{\mathcal{L}s\eta Q_{1}}{2}+\frac{\mathcal{L}s\eta Q_{1}\mathcal{T}^2\Phi}{2}+
            % &\frac{\mu(\eta\Vert{{Z}}\Vert^2 +2T^*\eta{Z}\sqrt{Q_2}+{T^*}^2\eta Q_2)}{s\Phi}
        \end{split}
    \end{align}
    Constant terms can be formulated as a quadratic function in terms of $\eta$. If the discriminant $\Delta\ge0$ holds, i.e., $\mathcal{T}^2s^3Q_2\Phi\ge2\mu W\mathcal{L}(1+\mathcal{T}^2s^2\Phi)$, there exists at least one learning rate $\eta$ for which the minimal gradient infinitely close to 0. The proof is completed.
    % Therefore, when 
    % \begin{align}
    %     \begin{split}
    %         \eta = \frac{-2s\Phi\mathcal{T}\sqrt{Q_1Q_2}}{\mathcal{L}s^2Q_1\Phi+\mathcal{L}s^2Q_1\mathcal{T}^2\Phi^2+2\mu Z^2+4\mu T^*Z\sqrt{Q_2}+2\mu{T^*}^2Q_2}
    %     \end{split}
    % \end{align}
    
\end{proof}

% \section{Additional experiments of security evaluation}
% \begin{figure}
%     \centering
%     \includegraphics[width = \linewidth]{figures/exp/reputation4.png}
%     \caption{Security evaluation among probability range (0-0.33).}
%     \label{fig:security}
% \end{figure}

\section{Additional experiments under replay attack}
\begin{figure}[ht]
    \centering
    \subfigure[100 devices]
    {
        \includegraphics[scale=0.14]{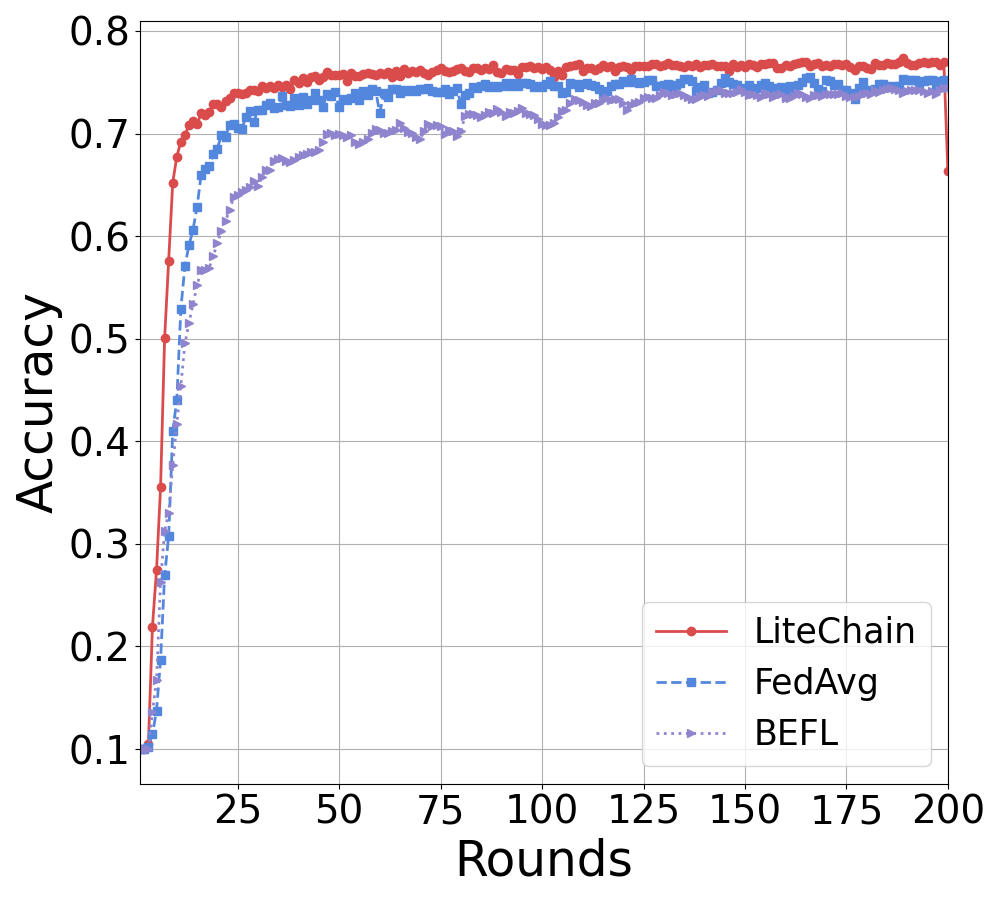}
        \label{a2_s100_5}
    }
    \subfigure[150 devices]
    {
        \includegraphics[scale=0.14]{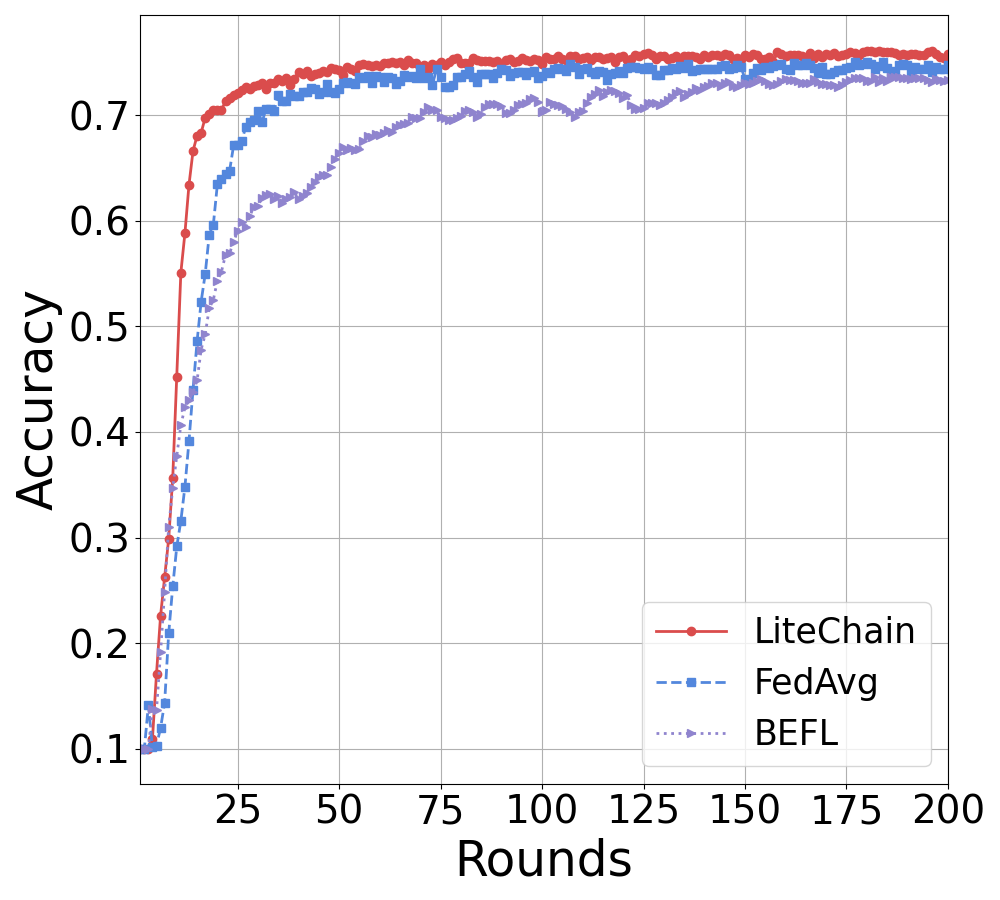}
        \label{a2_s150_5}
    }
    \subfigure[200 devices]
    {
        \includegraphics[scale=0.14]{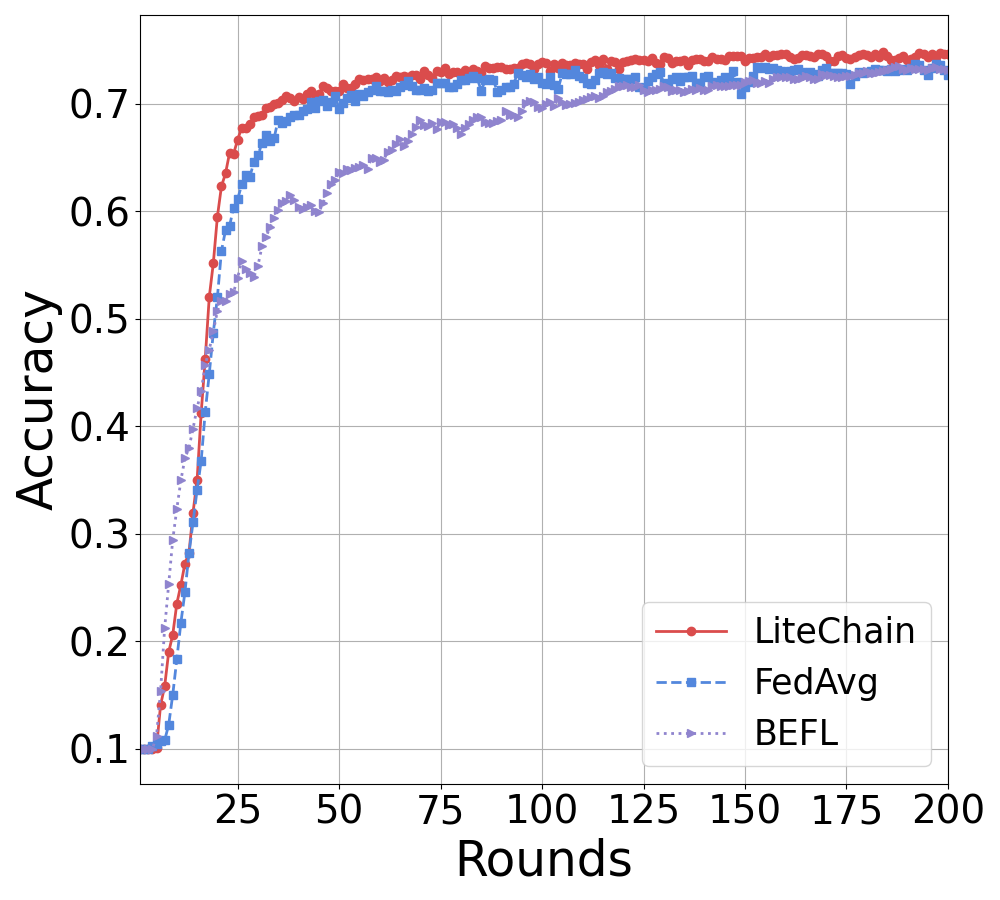}
        \label{a2_s200_5}
    }
    \subfigure[250 devices]
    {
        \includegraphics[scale=0.14]{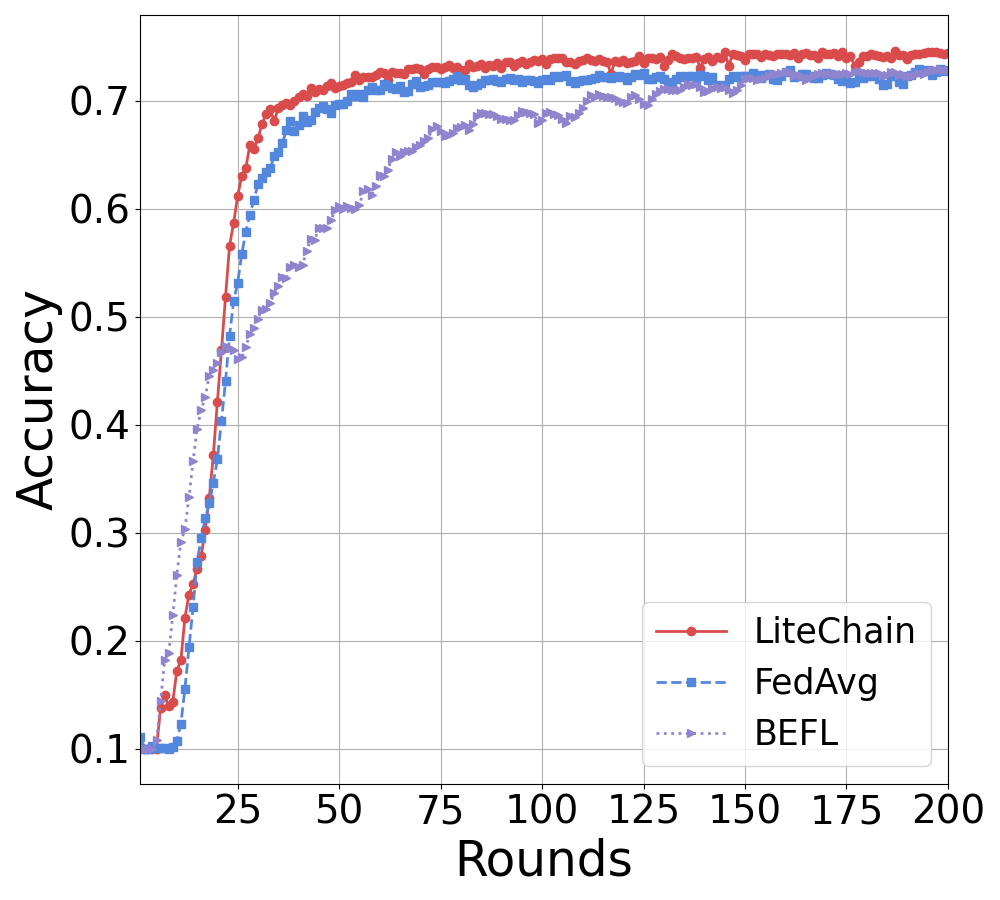}
        \label{a2_s250_5}
    }
    \label{exp:a2_iid_sup}
    \caption{Accuracy under replay attack on IID dataset (EFL represents BEFL without blockchain empowerment. FedAvg represents FLC without blockchain empowerment).}
\end{figure}

\begin{figure}[ht]
    \centering
    \subfigure[100 devices]
    {
        \includegraphics[scale=0.14]{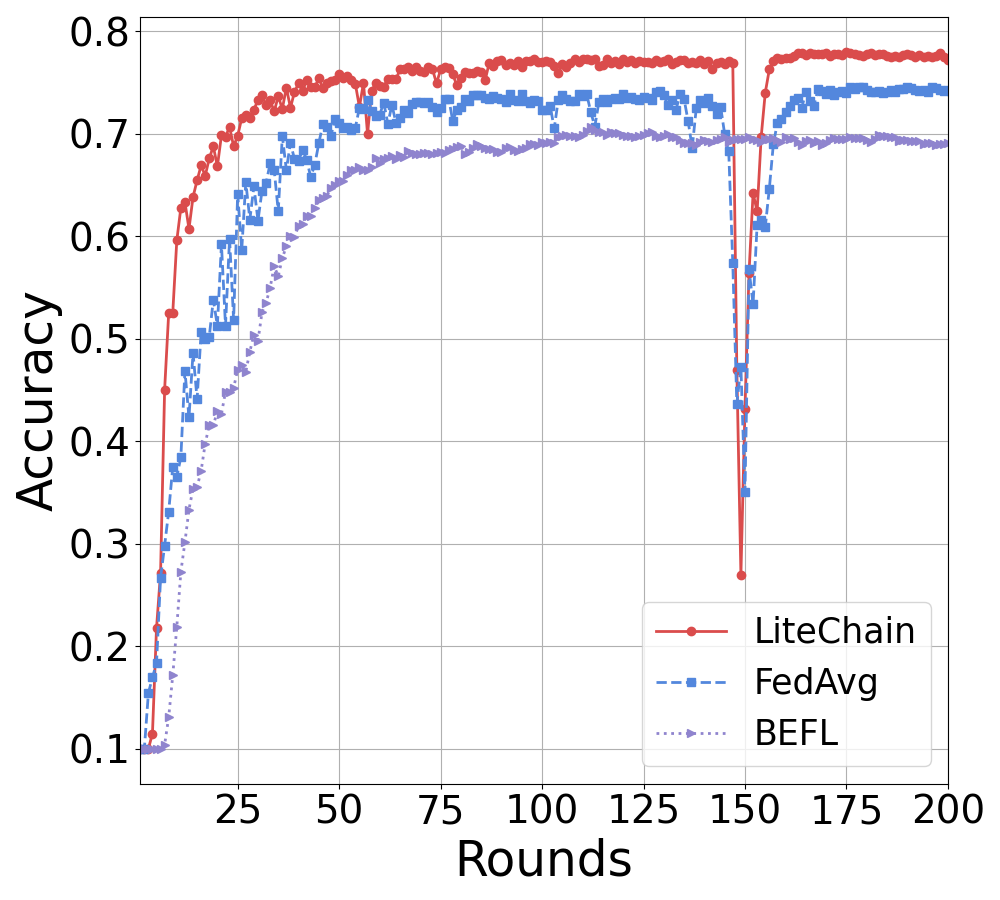}
        \label{a2_s100_02}
    }
    \subfigure[150 devices]
    {
        \includegraphics[scale=0.14]{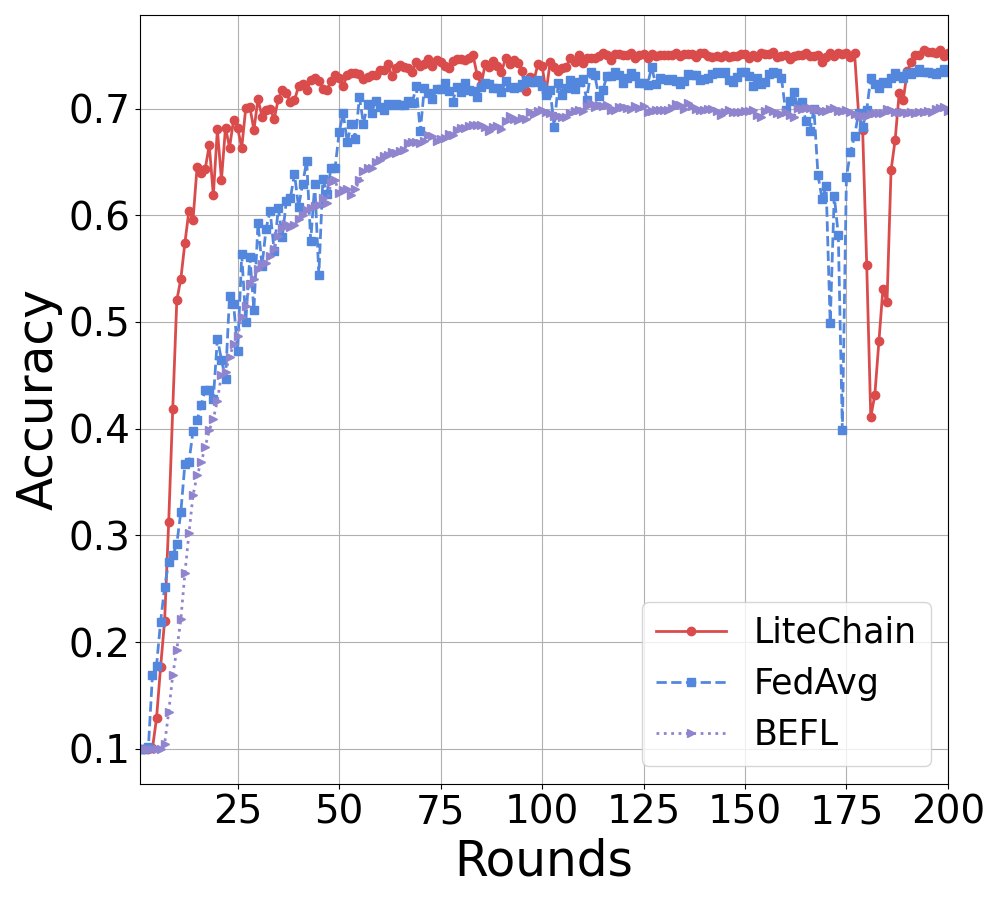}
        \label{a2_s150_02}
    }
    \subfigure[200 devices]
    {
        \includegraphics[scale=0.14]{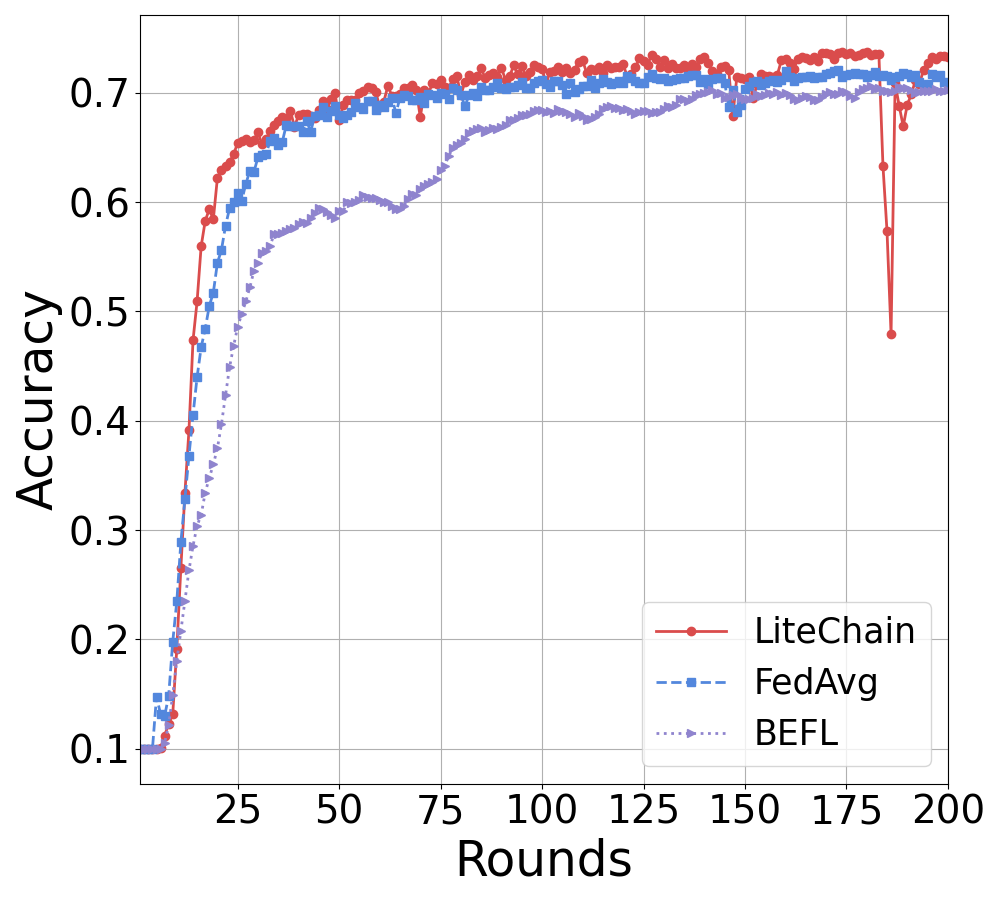}
        \label{a2_s200_02}
    }
    \subfigure[250 devices]
    {
        \includegraphics[scale=0.14]{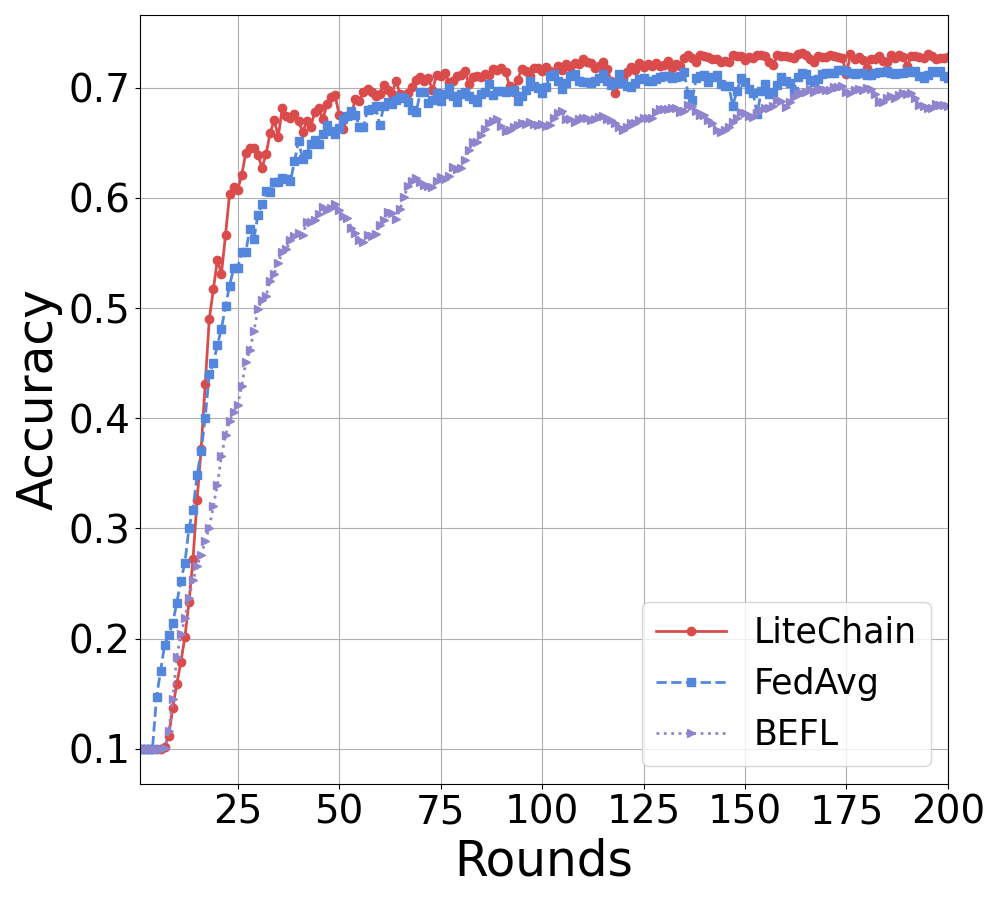}
        \label{a2_s250_02}
    }
    \label{exp:a2_niid_sup}
    \caption{Accuracy under replay attack on non-IID dataset (EFL represents BEFL without blockchain empowerment. FedAvg represents FLC without blockchain empowerment).}
\end{figure}

Figs. A-1 and A-2 evaluate the training performance under replay attackers across 100 devices to 250 devices with an attacking rate of 0.5 in IID and non-IID training datasets, respectively. For differentiation, we use EFL to denote BEFL without blockchain empowerment, and FedAvg to represent FLC without blockchain empowerment. 
The experimental outcomes clearly show that LiteChain outperforms FedAvg and EFL on performance across 100 to 250 devices. LiteChain achieves a higher convergence accuracy compared to other benchmarks. Its robustness is especially prominent as depicted in Fig. A-2 demonstrating its effectiveness in handling diverse training data distributions.

\end{appendices}
% \bibliographystyle{ieeetr}
% \bibliography{Refe}

\end{document}